\documentclass[%
 reprint,
 amsmath,amssymb,
 aps,
]{revtex4-2}

\usepackage{graphicx}
\usepackage{bm}
\usepackage{hyperref}

\usepackage{color}

\newcommand{\bnabla}{\bm{\nabla}}
\newcommand{\vs}{\bm{v}_\text{s}}
\newcommand{\vn}{\bm{v}_\text{n}}
\begin{document}

\title{Transition in vortex skyrmion structures in superfluid $^3$He-A driven by an analogue of the zero-charge effect}

\author{R. Rantanen}
\affiliation{%
 Department of Applied Physics, Aalto University, Finland
}%

\author{V.B. Eltsov}
\affiliation{%
 Department of Applied Physics, Aalto University, Finland
}%

\date{\today}

\begin{abstract}
In quantum electrodynamics, the zero-charge effect originates from the logarithmic dependence of the coupling constant in the action of the electromagnetic field on the ratio of the ultraviolet and infrared energy cutoffs. An analogue of this effect in Weyl superfluid $^3$He-A is the logarithmic divergence of the bending energy of the orbital anisotropy axis at low temperatures, where temperature plays the role of the infrared cutoff and the vector of the orbital anisotropy plays the role of the vector potential of the synthetic electromagnetic field for Weyl fermions. We calculate numerically the spatial distribution of the order parameter in rotating  $^3$He-A as a function of temperature. At temperatures close to the superfluid transition, we observe formation of vortex skyrmions known as the double-quantum vortex and the vortex sheet. These structures include alternating circular and hyperbolic merons as a bound pair or a chain, respectively. As temperature lowers towards absolute zero, we find a continuous transition in the vortex structures towards a state where the vorticity is distributed in thin tubes around the circular merons. For the vortex sheet, we present a phase diagram of the transition in the temperature -- angular velocity plane and calculations of the nuclear magnetic resonance response.
\end{abstract}

\maketitle


\section{\label{sec:introduction}Introduction}
Superfluidity of helium-3 is realized in the spin-triplet $p$-wave pairing state \cite{VollhardtWolfle}. The Cooper pairs have orbital momentum $L=1$ and spin $S=1$ and several distinct superfluid phases are found in the experiments \cite{LeggettReview}. The A phase, which is the focus of this work, is a chiral superfluid \cite{chiralsuperfluid}, where the components of a Cooper pair have equal spins, while all Cooper pairs have orbital momentum in the direction of the unit vector $\hat{\bm{l}}$. The gap $\Delta$ in the fermionic excitation spectrum of $^3$He-A is anisotropic and vanishes at two points on the Fermi surface along the orbital anisotropy axis defined by $\hat{\bm{l}}$. Near these gap nodes the Bogoliubov excitations have properties of Weyl fermions \cite{VolovikBook}.

Weyl nodes lead to several types of anomalous behaviour in $^3$He-A, including chiral anomaly \cite{bevannature,chiralanomaly,TCME}, thermal Nieh-Yan anomaly \cite{Nissinen2020}, Bogoliubov Fermi surface and non-thermal normal component in moving $^3$He-A \cite{zeroTnormalcomponent1,zeroTnormalcomponent2}, mass currents in the ground state \cite{anomalouscurrent,Nissinen2022}, and non-analytic coefficients in the expansion of free energy in terms of gradients of $\hat{\bm{l}}$ \cite{nonanalytic1,nonanalytic2}. For Weyl fermions in $^3$He-A, vector $\hat{\bm{l}}$ plays the role of the vector potential (up to a scaling factor), and thus its spatial variation and time dependence create a synthetic electromagnetic field. This effective electrodynamics possesses many features of the electrodynamics of quantum vacuum. In particular, the non-analyticity of the free energy is due to the logarithmic divergence of the coefficient $K_b\propto \ln\left(\Delta/T\right)$, associated with the term $[\hat{\bm{l}}\times(\bnabla\times\hat{\bm{l}})]^2$. This divergence is analogous to the running coupling constant in quantum electrodynamics \cite{VolovikBook}, with the gap $\Delta$ and the temperature $T$ playing the roles of the ultraviolet and infrared cut-offs, respectively \cite{zerocharge}. The logarithmically divergent running coupling constant in QED is due to the screening of electric charges by the polarized vacuum, known as the zero-charge effect.

In liquid $^3$He-A, the spatial distribution of $\hat{\bm{l}}$ is relatively flexible and can be manipulated by an external magnetic field, solid boundaries and rotation. The A phase tends to respond to rotation by creation of a continuous distribution of $\hat{\bm{l}}$ in the plane perpendicular to the rotation axis, formed from elements which carry both vortex and skyrmion topological charges, so-called vortex skyrmions. In this paper we present numerical calculations on continuous vortex skyrmion structures at low temperatures, where the divergence of the bending coefficient $K_b$ becomes relevant. The increased energy contribution from bending deformations of $\hat{\bm{l}}$ alters the vortex structures in a quantifiable manner. We have found a transition between two distinct core structures, and present a $\Omega$-$T$ phase diagram for the transition.

In Sec. \ref{sec:energy} we describe the different contributions to the free energy, and in Sec. \ref{sec:vortices} some possible realizations of vortex skyrmions in $^3$He-A: the double-quantum vortex and the vortex sheet. The numerical methods used to find the distribution of $\hat{\bm{l}}$ are described in Sec. \ref{sec:methods}. Section \ref{sec:nmr} briefly discusses the methods to calculate the nuclear magnetic resonance (NMR) response of the vortex skyrmion structures. The results of the paper are divided into four sections: in Sec. \ref{sec:modelvortex} we present calculations of a model ATC vortex and quantitative predictions on the low temperature structures, in Sec. \ref{sec:vortexsheet} results from vortex sheet calculations, in Sec. \ref{sec:doublequantumvortices} from separate double-quantum vortices and finally in Sec. \ref{sec:modelcomparison} we compare the results with the predictions made in Sec. \ref{sec:modelvortex}. The last section is dedicated to the conclusion.

\section{\label{sec:energy}Free energy density of $^3$H\lowercase{e}-A}
The order parameter in the A phase of superfluid $^3$He is separable in spin and orbital variables and has the form~\cite{VollhardtWolfle}
\begin{equation}
    A_{\mu j} = \Delta_A \hat{d}_\mu(\hat{m}_j + i\hat{n}_j)
    \label{eq:orderparameter}
\end{equation}
where unit vectors $\hat{\bm{m}}$, $\hat{\bm{n}}$ and $\hat{\bm{l}}$ form an orthonormal triad with $\hat{\bm{l}}$ being the direction of the orbital angular momentum of the Cooper pair, $\hat{\bm{d}}$ is a unit vector of spin anisotropy perpendicular to the preferred direction of the Cooper pair spin and $\Delta_A$ is the temperature- and pressure-dependent maximum gap in the energy spectrum of Bogoliubov quasiparticles.

The orientation of the order parameter anisotropy axes is affected by multiple competing interactions. The dipole interaction between magnetic momenta of the atoms forming the Cooper pair results in spin-orbit coupling, with the free energy density
\begin{equation}
    f_{\text{dip}} = \frac{3}{5}g_\text{d}\left[1-(\hat{\bm{l}}\cdot\hat{\bm{d}})^2\right].
    \label{eq:fdip}
\end{equation}
The spin-orbit energy is minimized when $\hat{\bm{l}}$ is parallel or antiparallel to $\hat{\bm{d}}$. The coefficient $g_d$ is expressed as \cite{VollhardtWolfle}
\begin{equation}
    g_\text{d}(T) = \frac{4}{3}\lambda_\text{d} N(0)\Delta_A(T)^2
    \label{eq:gd}
\end{equation}
where $\lambda_\text{d}\approx 5\times 10^{-7}$ is an approximately constant coupling parameter and $N(0)$ is the pressure-dependent density of states for one spin state.

In an external magnetic field $\bm{H}$, the spins of the Cooper pairs tend to align along it and thus $\hat{\bm{d}}$ favors orientation orthogonal to $\bm{H}$. The magnetic (Zeeman) energy density is
\begin{equation}
    f_{\text{mag}} = \frac{1}{2}\Delta\chi\left(\hat{\bm{d}}\cdot\bm{H}\right)^2.
    \label{eq:fmag}
\end{equation}

The coefficient $\Delta\chi$ is the difference between the two eigenvalues of the spin susceptibility tensor \cite{VollhardtWolfle}
\begin{equation}
    \Delta\chi = \frac{1}{2}\gamma^2\hbar^2 N(0)\frac{1-Y_0}{1+F_0^aY_0}
    \label{eq:deltachi}
\end{equation}
where $\gamma = -20 378 \text{ G}^{-1}\text{s}^{-1}$ is the gyromagnetic ratio of $^3$He, $Y_0$ is the Yosida function and $F_0^a \approx -0.756$ (at a pressure of $33\,\text{bar}$)\cite{F0a} is the pressure dependent spin-asymmetric Landau parameter.

Comparing Eqs. \eqref{eq:fmag} and \eqref{eq:fdip} one finds that the orientation effect of the magnetic field overcomes that of the spin-orbit interaction at the so-called dipolar field
\begin{equation}
H^* =  \left(\frac{6}{5}\frac{g_\text{d}}{\Delta\chi}\right)^{1/2}\approx 30\text{ G}.
\end{equation} 

In an isotropic superfluid such as $^4$He, the superfluid velocity $\vs$ is defined by the gradient of the phase $\phi$ of the order parameter $\psi = |\psi|e^{i\phi}$
\begin{equation}
    \vs^{(^4He)} = \frac{\hbar}{m_4}\bnabla\phi.
    \label{eq:vshe4}
\end{equation}
In superfluid $^3$He-A, with an anisotropic order parameter, the situation is more complicated. The order parameter in Eq. \eqref{eq:orderparameter} is invariant under relative gauge-orbit transformation and multiplying $A_{\mu j}$ with a phase factor $e^{i\phi}$ can be compensated by rotating $\hat{\bm{m}}$ and $\hat{\bm{n}}$ around $\hat{\bm{l}}$ by $-\phi$, ie. by transforming $\hat{\bm{m}} + i\hat{\bm{n}} \rightarrow e^{-i\phi}(\hat{\bm{m}}' + i\hat{\bm{n}}')$. The phase of the order parameter is then intrinsically linked to its orientation through the orbital angular momentum vector $\hat{\bm{l}}$. The superfluid velocity in the A phase is given by \cite{slowlyrotating}
\begin{equation}
    \vs = \frac{\hbar}{2m_3}\sum_i \hat{m}_i\bnabla\hat{n}_i
    \label{eq:vshe3}
\end{equation}
where $m_3$ is the mass of the $^3$He atom. Superflow is created by rotation of the orbital triad around a fixed $\hat{\bm{l}}$, as well as by changes in the orientation of $\hat{\bm{l}}$. The superfluid velocity in Eq. \eqref{eq:vshe3} is linked to the $\hat{\bm{l}}$ vector through the Mermin-Ho relation \cite{MerminHoRelation}:
\begin{equation}
    \bm{\omega} = \bnabla\times \vs = \frac{\hbar}{4m_3}\sum_{ijk}\epsilon_{ijk}\hat{l}_i\left(\bnabla\hat{l}_j\times\bnabla\hat{l}_k\right).
    \label{eq:merminhorelation}
\end{equation}
In the free energy we consider the terms with $\vs$ to be the kinetic energy. The kinetic energy density of $^3$He-A is
\begin{eqnarray}
    f_{\text{kin}} =\: &&\frac{1}{2}\rho_{\rm s\perp}\left(\hat{\bm{l}}\times\vs\right)^2 + \frac{1}{2}\rho_{\rm s\parallel}\left(\hat{\bm{l}}\cdot\vs\right)^2 \nonumber\\
    && +\: C\vs\cdot\left(\bnabla\times\hat{\bm{l}}\right) - C_0\left(\vs\cdot\hat{\bm{l}}\right)\hat{\bm{l}}\cdot\left(\bnabla\times\hat{\bm{l}}\right)
    \label{eq:fkin}
\end{eqnarray}
where $\rho_{\rm s\perp}$ and $\rho_{\rm s\parallel}$ are the superfluid density in the direction perpendicular and parallel to $\hat{\bm{l}}$, respectively, and $C$ and $C_0$ are coefficients related to the superflow. The first two terms in Eq. \eqref{eq:fkin} can be written as
\[\frac{1}{2}\rho_{\rm s\perp}\vs^2 - \frac{1}{2}\left(\rho_{\rm s\perp} - \rho_{\rm s\parallel}\right)\left(\hat{\bm{l}}\cdot\vs\right)^2.\]
It is seen from here that it is energetically favorable for $\hat{\bm{l}}$ to be oriented along the superfluid velocity $\vs$, since $\rho_{\rm s\perp} > \rho_{\rm s\parallel}$ as the gap in the quasiparticle energy spectrum is at maximum in the direction perpendicular to $\hat{\bm{l}}$ while it is zero along $\hat{\bm{l}}$.

In a superfluid rotating with constant angular velocity $\bm{\Omega}$, the normal fluid rotates as a solid body with the velocity
\begin{equation}
    \vn = \bm{\Omega}\times\bm{r}.
    \label{eq:vn}
\end{equation}
For a rotating system, two extra terms are included in the free energy of the whole fluid, $\frac{1}{2}\rho_{\rm n}\vn^2$ and $-\bm{\Omega}\cdot\bm{L}$ where $\bm{L}$ is the total angular momentum. Adding these terms is equivalent to transforming $\vs\rightarrow \vs-\vn$ in Eq. \eqref{eq:fkin}, when a constant term $-\frac{1}{2}\rho\vn^2$ is omitted.

In the absence of an external magnetic field and rotation, the minimum energy configuration in the bulk corresponds to the uniform order parameter. This is due to the elastic energy associated with changes in the orientation of the $\hat{\bm{l}}$ and $\hat{\bm{d}}$ vectors
\begin{widetext}
\begin{equation}
    f_{\rm el} = \frac{1}{2}K_s\left(\bnabla\cdot\hat{\bm{l}}\right)^2 + \frac{1}{2}K_t\left[\hat{\bm{l}}\cdot\left(\bnabla\times\hat{\bm{l}}\right)\right]^2 + \frac{1}{2}K_b\left[\hat{\bm{l}}\times\left(\bnabla\times\hat{\bm{l}}\right)\right]^2 + \frac{1}{2}K_5\sum_a\left[\left(\hat{\bm{l}}\cdot\bnabla\right)\hat{d}_a\right]^2 + \frac{1}{2}K_6\sum_a\left(\hat{\bm{l}}\times\bnabla\hat{d}_a\right)^2
    \label{eq:fel}
\end{equation}
\end{widetext}
where the terms with coefficients $K_s$, $K_t$ and $K_b$ correspond to splay-, twist-, and bend-like deformations in the $\hat{\bm{l}}$-vector texture, respectively. The terms with $K_5$ and $K_6$ are related to changes in the $\hat{\bm{d}}$ vector orientation.

The temperature dependent coefficients $K_i$ in the elastic energy are calculated using Cross's weak coupling gas model \cite{nonanalytic2}, following Fetter \cite{FetterCoeff} and using the Cross functions calculated by Thuneberg \cite{ThunebergCoeff}. The coefficients entering in the free energy density are presented in Appendix \ref{sec:coefficients}. The bending coefficient $K_b$ warrants special attention, as it is connected to the zero-charge effect in $^3$He-A \cite{VolovikBook}. It is logarithmically divergent as $K_b(T) = K_{b1} + K_{b2} \ln (T_{\rm c}/T)$ when $T\to 0$ owing to nodes in the energy gap in the spectrum of Bogoliubov quasiparticles \cite{nonanalytic2}.

The boundary conditions imposed by the container walls are also crucial in determining the texture. Most importantly, the $\hat{\bm{l}}$ vectors are forced perpendicular to the boundary surface \cite{lboundarycondition}. This means that the $\hat{\bm{l}}$ texture cannot be uniform in a system with finite size. In addition, the superflow through the walls must be zero \cite{brinkmancross}, meaning that in the rotating frame $\vs - \vn$ must be aligned parallel to the surface at the boundaries. Ignoring magnetic relaxation effects near the surface, the spin currents and thus the gradients of the spin anisotropy vector components $\bnabla\hat{d}_a$ are aligned parallel to the boundaries.

\section{\label{sec:vortices}Continuous vortices}
The form of superfluid velocity in $^3$He-A, Eq. \eqref{eq:vshe3}, allows for the formation of vortex structures that do not require the suppression of the amplitude of the order parameter like in conventional superfluids and superconductors. As shown by the Mermin-Ho relation \eqref{eq:merminhorelation}, the vorticity $\bm{\omega}$ can be non-zero in regions where $\hat{\bm{l}}$ is non-uniform. This means that non-singular vortices with continuous vorticity can exist in the superfluid.

In this paper, we focus on continuous vortex structures. In $^3$He-A hard-core defects where the order parameter is suppressed are also possible. These types of structures are generally not formed when rotation is started in the superfluid state, due to their high critical velocity of nucleation compared to continuous vortices \cite{criticalvelocity,vortexphasediagram}. 

On a closed path around a vortex, the circulation is given by \cite{quantizedvorticity}
\begin{equation}
    \nu\kappa_0 = \oint \bm{v_s}\cdot d\bm{r} = \frac{\hbar}{2m_3}\mathcal{S}(\hat{\bm{l}})
    \label{eq:circulation}
\end{equation}
where $\kappa_0 = h/2m_3$ is the quantum of circulation for $^3$He, $\nu$ is the number of circulation quanta and $\mathcal{S}(\hat{\bm{l}})$ is defined as the area on the unit sphere covered by the orientations of $\hat{\bm{l}}$ inside the domain bounded by the closed path. In Eq.~\eqref{eq:circulation}, the first integral along the path is the usual expression of the topological invariant defining quantized vortices. The second integral over the area enclosed by the path is the topological invariant usually used for defining skyrmions. The equivalence of the two expressions follows from the Mermin-Ho relation \eqref{eq:merminhorelation}. The continuous vortex structures in $^3$He-A, surrounded by the volume where $\hat{\bm{l}}$ lies in a plane,  possess both invariants, that is, they are simultaneously quantized vortices and skyrmions. The in-plane orientation of $\hat{\bm{l}}$ in the external regions, needed to ensure integer values of integrals in Eq.~\eqref{eq:circulation}, is usually provided by the boundary conditions at the sample walls or by the combination of the spin-orbit \eqref{eq:fdip} and Zeeman \eqref{eq:fmag} interactions in the applied magnetic field.

The simplest continuous vortex structure contains one quantum of circulation, and is known as the Mermin-Ho vortex. In the core of a Mermin-Ho vortex, the $\hat{\bm{l}}$-vectors rotate out of the plane, covering exactly half of a unit sphere. Single Mermin-Ho vortices are observed in narrow cylinders \cite{TakagiNMR} where they are stabilized by the effect of the container walls on the orientation of $\hat{\bm{l}}$.

\begin{figure}
    \centering
    \includegraphics[width=\linewidth,keepaspectratio]{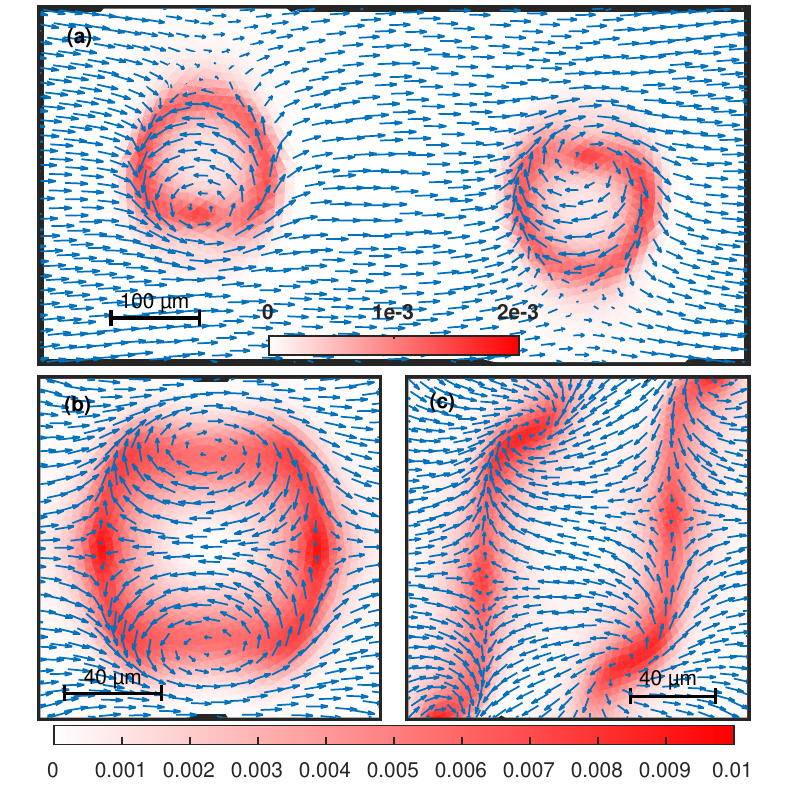}
    \caption{Three different continous vortex structures with four quanta of circulation in $^3$He-A. The blue arrows represent the $\hat{\bm{l}}$ vector texture and the background color is vorticity in units of $\kappa/\mu \text{m}^2$. \textbf{(a)} Two double-quantum vortices at $\Omega = 0.30\,\text{rad/s}$. \textbf{(b)} A circular vortex sheet at $\Omega = 5.70\,\text{rad/s}$. \textbf{(c)} Two separate vortex sheets, each with two quanta of circulation at $\Omega = 5.70\,\text{rad/s}$. The sheets are connected to the container walls outside the shown area. }
    \label{fig:vortexstructures}
\end{figure}

A skyrmion in $^3$He-A is a topological object where the $\hat{\bm{l}}$ vectors cover the whole unit sphere, with $\nu = 2$ quanta of circulation. An axisymmetric skyrmion known as the Anderson-Toulouse-Chechetkin (ATC) vortex \cite{ATC1,ATC2} is the simplest model for a double-quantum vortex in $^3$He-A. The ATC vortex with the axis along $\hat{\bm{z}}$ consists of a topological soliton separating a core region with $\hat{\bm{l}}=\hat{\bm{z}}$ from an outer region with $\hat{\bm{l}}=-\hat{\bm{z}}$. In a finite system, however, the axisymmetry of the structure is broken by the bulk $\hat{\bm{l}}$ texture, which is confined to the $xy$ plane  by the boundary conditions at the walls parallel to $\hat{\bm{z}}$. The non-axisymmetric double-quantum vortex, shown in figure \ref{fig:vortexstructures}a, consists of two merons: one circular and one hyperbolic. The circular meron covers the top half of the unit sphere and the hyperbolic meron the lower half. The vorticity $\bm{\omega}$ in a double-quantum vortex (DQV) is concentrated in a cylindrical tube around the axis of the vortex, with a vorticity free region between the two merons \cite{ThunebergPNAS}. These structures typically appear in systems with magnetic fields above the dipolar field \cite{vortexphasediagram}, where the order parameter is "dipole-unlocked" inside the core, so that $\hat{\bm{d}}$ is forced to stay in-plane by the magnetic field while $\hat{\bm{l}}$ covers all possible directions. In lower magnetic fields, the merons arrange into a square lattice where a cell consists of two hyperbolic and two circular vortices, totaling four circulation quanta. 

At high rotation velocities, the vortex sheet is the preferred texture over separate vortex lines \cite{vortexphasediagram}. A vortex sheet is a chain of alternating circular and hyperbolic merons confined inside a topological soliton that separates two regions minimizing spin-orbit interaction energy with one having $\hat{\bm{l}} \uparrow\uparrow \hat{\bm{d}}$ and the other $\hat{\bm{l}} \uparrow \downarrow \hat{\bm{d}}$. A circular vortex sheet texture is shown in figure \ref{fig:vortexstructures}b. The sheet can be connected to the container walls (figure \ref{fig:vortexstructures}c), and as the rotation speed is increased, vortices enter the system through these connection points and the vortex sheet begins to spiral, meandering around the volume while keeping the soliton walls equidistant \cite{topologicaldefects}. After a wall-connected soliton has appeared in the system, it becomes difficult to nucleate separate vortices, as the critical velocity of formation of new merons at the connection of the vortex sheet to the wall is lower than that of separate DQVs \cite{SheetMeasurements,ThunebergSheet}. 

\section{\label{sec:methods}Numerical methods}
We find the lowest-energy state of $^3$He-A in the London limit through numerical minimization. The calculation is done in two dimensions and we assume that the system is uniform in the $z$ direction. The function to be minimized is the total energy per unit height $F$, defined as
\begin{equation}
    F = \int_S \left(f_{\rm dip} + f_{\rm mag} + f_{\rm kin} + f_{\rm el}\right) dS
    \label{eq:ftotal}
\end{equation}
The minimization is performed with respect to the spin anisotropy vector $\hat{\bm{d}}$ and the orbital triad consisting of the three orthonormal vectors $\hat{\bm{l}}$, $\hat{\bm{m}}$ and $\hat{\bm{n}}$. The parameterization of the triad is done using unit quaternions. Quaternions have the benefit of reducing the number of variables from nine to only four, while also avoiding the problems associated with Euler angles like singularities and gimbal lock. The $\hat{\bm{d}}$ vector is parameterized with azimuthal and polar angles $\alpha$ and $\beta$, where these problems are avoided by choosing the polar axis along the magnetic field direction. The magnetic energy \eqref{eq:fmag} ensures that the polar angle should always be nonzero during the minimization process in the dipole-unlocked regime we are interested in. The parameterization is presented in more detail in the Appendix.

The calculations are done on two-dimensional circular domains with varying radii, which are meshed into triangular elements. The resolution used is limited by the available computational time and memory and varies between $3.5\,\mu\mathrm{m}$ and $10\,\mu\mathrm{m}$ depending on the size of the system in question. The integration in Eq. \eqref{eq:ftotal} is done using Gaussian quadrature rules on the triangles. MATLAB is used to find the texture that minimizes the total energy using the BFGS method. The boundary conditions are implemented with the barrier method by adding an additional energy term that penalizes parameter values that would violate the boundary conditions.

The coefficients for the free energy terms are calculated at a pressure of 33 bar and varying temperatures. The magnetic field in the simulations is set to  0.55\,T, as that is a high enough value for $^3$He-A to be stable down to zero temperature.

\section{\label{sec:nmr}Nuclear magnetic resonance}
Nuclear magnetic resonance (NMR) is a useful experimental tool for studies of superfluid helium-3. Different order parameter structures can usually be distinguished from the NMR absorption spectrum. In $^3$He-A, the long-range order of $\hat{\bm{l}}$ and $\hat{\bm{d}}$ together with the spin-orbit interaction leads to spontaneously broken spin-orbit symmetry \cite{SBSOS1,*SBSOS2}. The coupling between the spin and orbital degrees of freedom leads to an extra torque applied to the precessing spin in NMR experiments which allows us to probe the $\hat{\bm{l}}$ texture.  Different vortex structures result in satellites in the NMR spectrum with characteristic frequency shifts \cite{NMRsignatures}. We consider longitudinal NMR here, because at low temperatures we are interested in, $^3$He-A in bulk is stable at relatively high magnetic fields and the longitudinal resonance frequencies are independent of the magnetic field strength.

Assuming a static equilibrium texture for $\hat{\bm{d}}=\hat{\bm{d}}_0$, we parametrize the deviation of $\hat{\bm{d}}$ from the equilibrium due to the oscillating field with two parameters $d_H$ and $d_\theta$ for the deviation along the field and perpendicular to the field, respectively. The $\hat{\bm{d}}$ vector in the presence of the oscillating field is
\begin{equation}
    \hat{\bm{d}} = \hat{\bm{d}}_0 + (\hat{\bm{H}}\times\hat{\bm{d}}_0)d_\theta + \hat{\bm{H}}d_H
\end{equation}
where $\hat{\bm{H}}$ is a unit vector in the direction of the static magnetic field. The longitudinal NMR resonance frequencies are given by the Schrödinger-type equation \cite{ThunebergNMR}
\begin{equation}
    (\mathcal{D} + U_\parallel)d_\theta = \alpha_\parallel d_\theta
    \label{eq:nmreigenvalue}
\end{equation}
where the operator $\mathcal{D}$ is defined as
\begin{equation}
    \mathcal{D}f = -\frac{5}{6}\frac{K_6}{g_d}\bnabla^2 f - \frac{5}{6}\frac{K_5-K_6}{g_d}\bnabla\cdot\left[\hat{\bm{l}}(\hat{\bm{l}}\cdot\bnabla)f\right]
    \label{eq:operatorD}
\end{equation}
and the potential is
\begin{equation}
    U_\parallel = 1 - (\hat{\bm{H}}\cdot\hat{\bm{l}})^2 - 2[\hat{\bm{H}}\cdot(\hat{\bm{l}}\times\hat{\bm{d}})]^2
    \label{eq:nmrpotential}
\end{equation}
The resonance frequencies are related to the eigenvalues $\alpha_\parallel$ in \eqref{eq:nmreigenvalue} by
\begin{equation}
    \omega_\parallel^2 = \Omega_{\rm A}^2\alpha_\parallel
    \label{eq:resonancefrequencies}
\end{equation}
where $\Omega_{\rm A}$ is the Leggett frequency of the A phase.

The NMR resonance frequencies are calculated by solving the eigenvalue problem \eqref{eq:nmreigenvalue} using the finite element method (FEM). The same mesh from the energy minimization is used and the equation is discretized using linear shape functions. The calculated eigenfunctions $\psi_k$ are normalized so that
\begin{equation}
    \int_S |\psi_k|^2 dS = 1
    \label{eq:eigenfunctionnormalization}
\end{equation}
for each eigenfunction. The convenience of FEM is that the method automatically enforces the Neumann boundary conditions for the spin waves. Dissipation effects are not taken into account in the NMR calculation, which may affect the results quantitatively.

\section{\label{sec:modelvortex}ATC Vortex}

\begin{figure*}
    \centering
    \includegraphics[width=\linewidth,keepaspectratio]{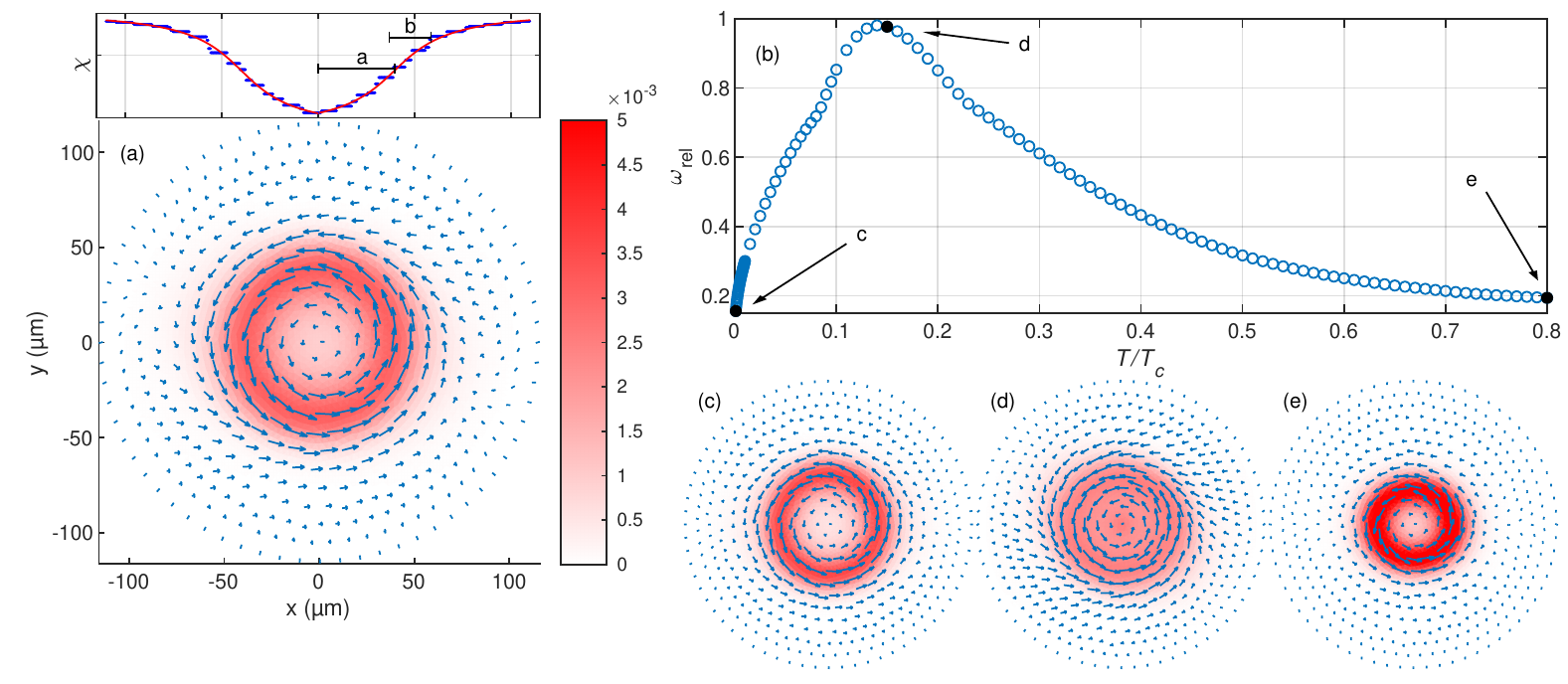}
    \caption{\label{fig:modeltexture} \textbf{(a)} Spatial distribution (texture) of the orbital anisotropy axis $\hat{\bm{l}}$ (blue arrows) in the ATC vortex, where $\hat{\bm{l}}$ is parallel to the $\hat{\bm{z}}$-axis in the center of the computational domain and antiparallel to it at the boundary. The color indicates the distribution of vorticity in units of $\kappa/\mu \text{m}^2$. The predicted vorticity tube is clearly visible. (Inset) An example of a fit (line) of the angle $\chi$ determined from the simulation results along $y=0$ (points) to Eq.\eqref{eq:chi}. The radius $a$ and thickness $b$ are determined from the fit and illustrated in the figure. The calculations are performed at $T=0.01T_{\rm c}$ and $\Omega = 2.3$ rad/s. \textbf{(b)} The relative vorticity $\omega_{\rm rel}$ of the ATC vortex as a function of temperature. The values are calculated from the temperature sweep. \textbf{(c)}-\textbf{(e)} ATC vortex texture and vorticity distribution at temperatures $0.001T_{\rm c}$, $0.15T_{\rm c}$ and $0.80T_{\rm c}$, respectively. The corresponding points are marked with filled black circles in (b). The color scale is the same as in (a).}
\end{figure*}

At low temperatures, the logarithmic divergence of the $K_b$ coefficient implies that the manifested structure should be one that minimizes bending deformations. Since the $\hat{\bm{l}}$ vectors in a double-quantum vortex cover the whole unit sphere, this type of deformation cannot be avoided completely. In a circular ATC vortex, the $\hat{\bm{l}}$ vectors pointing along $\hat{\bm{z}}$ in the center are separated from an external region with $\hat{\bm{l}}$ along $-\hat{\bm{z}}$ by a topological twist soliton. There are two defining lengths for the soliton: the radius $a$ and the thickness $b$, illustrated in the inset in figure \ref{fig:modeltexture}a. Along the radial direction there is only twist deformation, but on a loop around the vortex center $\hat{\bm{l}}$ bends (and splays) a full rotation, so that the elastic energy can be estimated as $F_{el} \sim K_b(b/a) + K_t(a/b)$. Therefore we expect that as the temperature decreases, the radius of the vortex increases, while the thickness of the soliton decreases. According to the Mermin-Ho relation \eqref{eq:merminhorelation}, the vorticity $\bm{\omega}$ is concentrated in the soliton, with no vorticity in the relatively uniform center.

The structure suggested by Volovik \cite{volovik} is this type of ATC vortex, where the $\hat{\bm{l}}$ texture in the soliton is
\begin{equation}
    \hat{\bm{l}} = \cos\chi(r)\hat{\bm{z}} + \sin\chi(r)\hat{\bm{\varphi}}
    \label{eq:domainwall}
\end{equation}
\begin{equation}
    \chi(r) = \text{arccot}\left(-\frac{r-a}{b}\right) \label{eq:chi}
\end{equation}
where $r$, $\varphi$ and $z$ are the cylindrical coordinates. Following the derivation by Volovik, but including the effect of the rotation of the system in the kinetic energy gives us the following formulas for the soliton radius and thickness:
\begin{equation}
    a = \left[\frac{\rho\left(\frac{\hbar}{m_3}\right)^2}{2\rho\Omega\frac{\hbar}{m_3} + \frac{12}{5}\pi g_d\left(\frac{K_t}{K_b}\right)^{1/2}}\right]^{1/2}
    \approx \left(\frac{\hbar}{2m_3\Omega}\right)^{1/2}
    \label{eq:a}
\end{equation}
\begin{equation}
    b = a\left(\frac{K_t}{K_b}\right)^{1/2}
    \label{eq:b}
\end{equation}
The magnetic field $\bm{H}$ is transverse to the cylinder axis. To simplify the derivation, $\hat{\bm{d}}$ is assumed to be uniformly along $\hat{\bm{z}}$ and the terms with $C$ and $C_0$ in Eq. \eqref{eq:fkin} have been ignored. At higher rotation speeds, however, these terms turn out to have a considerable effect.

At finite rotation speeds, the temperature dependence of $a$ is negligible, which gives the $K_b$-independent expression in Eq.~\eqref{eq:a}. Therefore the effects of the logarithmic divergence of $K_b$ should only be seen in the narrowing of the domain wall.

We numerically calculate the structure of the ATC vortex in a circular domain with a radius $R = 115\,\mu\mathrm{m}$. As the initial configuration for the energy minimization we set $\hat{\bm{l}}$ parallel to $\hat{\bm{z}}$ at the center and antiparallel to $\hat{\bm{z}}$ at the boundary, with a linear rotation around the radial direction in between. The boundary condition is applied such that $\hat{\bm{l}}$ stays antiparallel to the $z$-axis at the edge of the calculation domain. The magnetic field direction is along the $y$-axis and accordingly the $\hat{\bm{d}}$ vectors are uniformly pointing in the $z$ direction. The temperature is set to $T = 0.005T_{\rm c}$. The angular velocity is initially set to the value $\Omega = 2.7\,\text{rad/s}$, as minimization with zero rotation results in the vortex drifting to the walls of the simulation box. The result of this minimization is then used as an initial state for the angular velocity sweep. An example of the ATC vortex texture is presented in figure \ref{fig:modeltexture}a.

\begin{figure*}
\includegraphics[width=\linewidth,keepaspectratio]{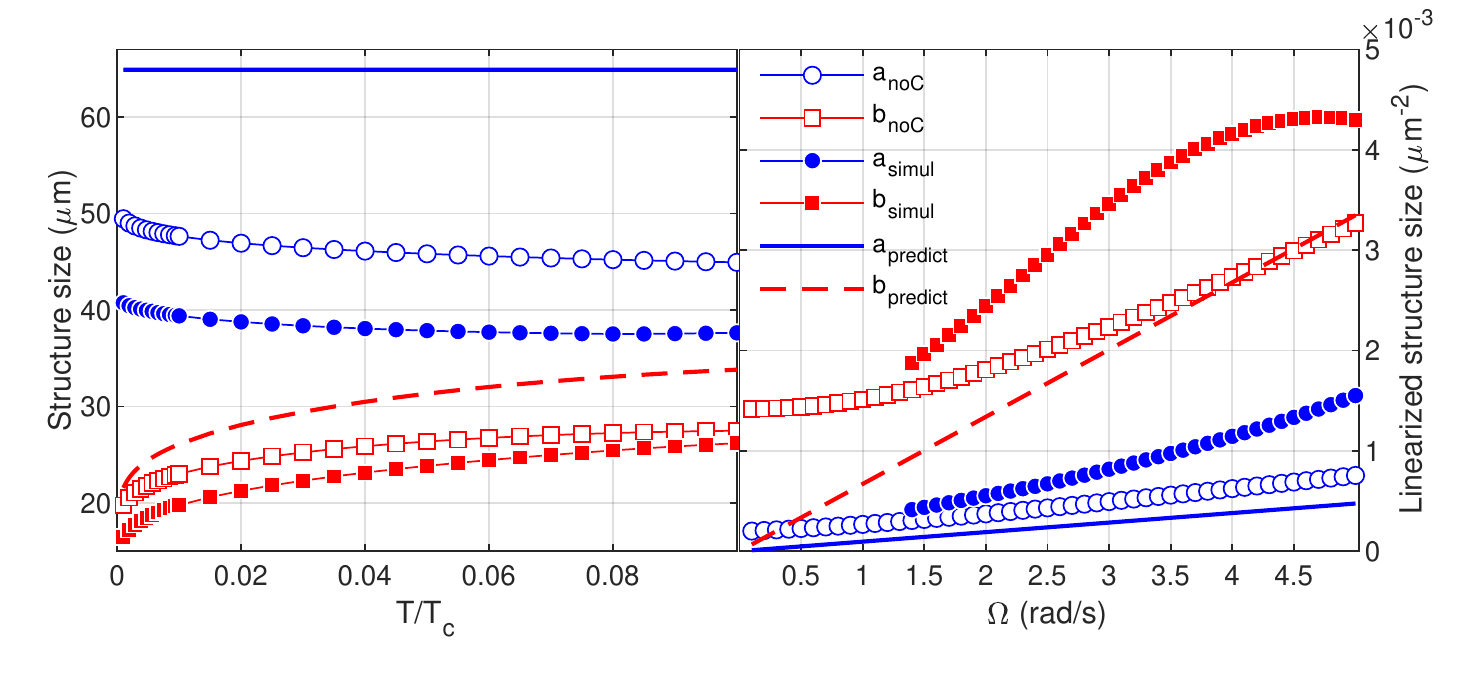}
\caption{\label{fig:modelparameters} Size of the vorticity tube in the ATC vortex. (\textit{Left}) The width and radius of the vorticity tube in the ATC vortex, figure \ref{fig:modeltexture}, as a function of temperature at $\Omega = 2.3$ rad/s. The solid blue line and dashed red line correspond to the values of $a$ and $b$ calculated from Eqs.~\eqref{eq:a} and \eqref{eq:b}, respectively, and the filled blue circles ($\bullet$) and red squares ($\blacksquare$) are the corresponding values determined from the simulation results. The results from a simulation run where the kinetic energy terms with $C$ and $C_0$ coefficients have been set to zero are shown as empty blue circles ($\circ$) and red squares ($\square$) for $a$ and $b$, respectively. The temperature dependence is similar in the numerical calculations and the model, although the radius is predicted to be significantly larger than numerically calculated.
(\textit{Right}) $a^{-2}$ and $b^{-2}$ as functions of rotation speed. From the equations \eqref{eq:a} and \eqref{eq:b}, the behaviour is expected to be linear in these coordinates. The symbols are the same as in the left panel. The omission of the $C$ terms provides a much better match for $\Omega$-dependence. At low angular velocities, the discrepancy grows as the size of the computation domain limits the size of the vortex.}
\end{figure*}

The value of the angular velocity $\Omega$ is gradually decreased in steps of $0.1$ rad/s, starting from the initial value $\Omega = 2.7$ rad/s. At each step of $\Omega$, the previous minimization result is used as the new initial state, in order to mimic a realistic continuous deceleration. An increasing $\Omega$ sweep is also performed, similarly starting at the initial $\Omega = 2.7$ rad/s.

The lowest energy during the $\Omega$ sweep is achieved at $\Omega = 2.3$ rad/s. This value for $\Omega$ is used in the temperature sweep. The temperature sweep is started from $T/T_{\rm c} = 0.001$ and the temperature is increased gradually in steps, ending at a temperature of $T/T_{\rm c} = 0.8$. More points are calculated at lower temperatures $T/T_{\rm c} < 0.1$, as that is the region where the logarithmic divergence of the $K_b$ coefficient is relevant.

From the simulation results we find the radius and width of the topological soliton by fitting values of $\cos^{-1}(\hat{l}_z)$ to the model $\chi(r)$ dependence in Eq.~\eqref{eq:chi}, with $a$ and $b$ as the fitting parameters. An example fit is shown in the inset in figure \ref{fig:modeltexture}a. The fits are done along multiple radial lines going around the whole simulation disk, and the radius and width values $a$ and $b$ are taken as the mean of the fitting parameters over each line. There is slight axial asymmetry in the texture, due to the dipole interaction in the transverse field.

The numerically calculated and predicted values of $a$ and $b$ for the ATC vortex are shown in figure \ref{fig:modelparameters}. The predicted dependence $a^{-2} \propto \Omega$ and $b^{-2} \propto \Omega$ is not so clear in the simulation results. A possible reason is the omission of the $C$ and $C_0$ terms in the kinetic energy in the model derivation. To test this possibility, a simulation with the same setup but without these terms has been performed, the results of which are also shown in figure~\ref{fig:modelparameters}. 

The analytic model is indeed in closer agreement with the simulation without $C$ terms, especially at higher angular velocities. At low rotation speeds, the agreement becomes worse, as the numerically calculated structure becomes limited by the simulation domain.

\begin{figure*}
    \centering
    \includegraphics[width=\linewidth,keepaspectratio]{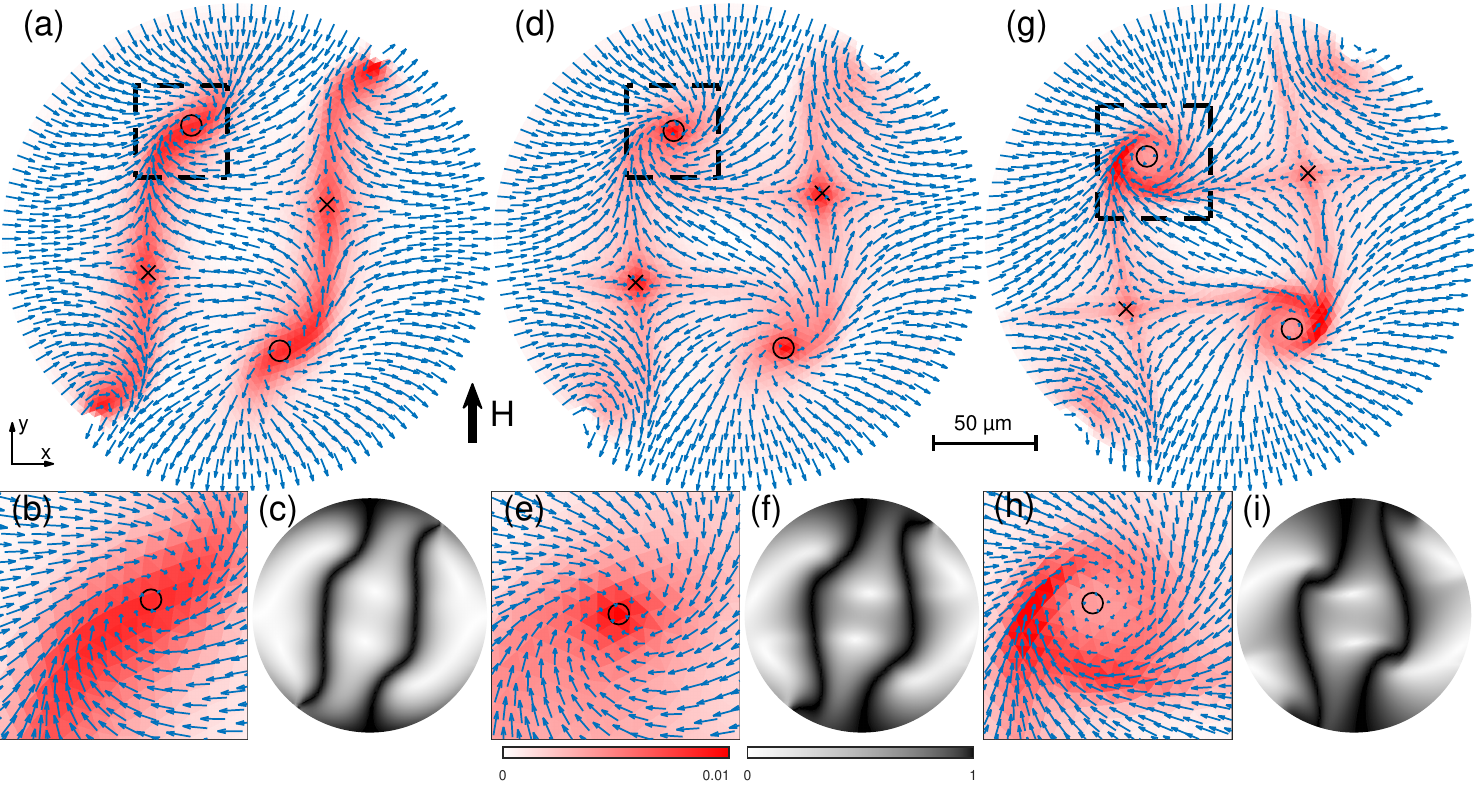}
    \caption{\label{fig:zerochargetransition}The zero-charge transition in the vortex sheet. The texture in (a-c) is calculated at $0.8T_{\rm c}$, the one in (d-f) at $0.2T_{\rm c}$ and the one in (g-i) at $0.006T_{\rm c}$. \textbf{(a)} The $\hat{\bm{l}}$-vector texture, shown using blue arrows, with the vorticity distribution given by the colormap in units of $\kappa/\mu\text{m}^2$. The two vortex sheets are clearly visible, with vorticity distributed uniformly along the sheet. The centers of the circular and hyperbolic merons are marked with circles and crosses, respectively. \textbf{(b)} A close-up of the circular meron area marked by the dashed line in (a). The point where $\hat{l}_z=1$ is marked with a circle. \textbf{(c)} A density plot for the value of $|\hat{\bm{l}}\times\hat{\bm{d}}|$. The regions where $\hat{\bm{l}}$ and $\hat{\bm{d}}$ are not aligned are darker. \textbf{(d)} The texture at $0.2T_{\rm c}$. The vorticity in the sheets has concentrated near the merons. \textbf{(e)} Close-up of the circular meron marked with a dashed line in (d). \textbf{(f)} The $|\hat{\bm{l}}\times\hat{\bm{d}}|$ density plot at $0.2T_{\rm c}$. The background soliton of the sheets is still there despite the change in vorticity distribution. \textbf{(g)} The texture at $0.006T_{\rm c}$, after the transition. The vorticity near the circular merons has formed the distinct tube shapes associated with the zero-charge effect. \textbf{(h)} A close-up of the circular meron marked with a dashed line in (g). \textbf{(i)} The $|\hat{\bm{l}}\times\hat{\bm{d}}|$ density plot at $0.006T_{\rm c}$. The two sheets persist after the transition.}
\end{figure*}

During the increasing temperature sweep, the vorticity distribution in the vortex becomes more uniform. After becoming completely uniform at around $T = 0.15T_{\rm c}$, the tube distribution reappears, as shown in figures \ref{fig:modeltexture}c-d. A good quantitative indicator for the presence of these vorticity tubes is found to be the relative vorticity $\omega_{\rm rel}$ defined as the ratio of the vorticity at the center of the vortex (where $\hat{l}_z=1$) to the maximum vorticity in the system
\begin{equation}
    \omega_{\rm rel} = \frac{|\bm{\omega}|}{\max(|\bm{\omega}|)}.
\end{equation}
The relative vorticity for the ATC vortex is plotted in figure \ref{fig:modeltexture}b. At both low and high temperatures $\omega_{\rm rel}$ approaches zero, but the effect is caused by different interactions. At temperatures below $T = 0.15T_{\rm c}$, the $\hat{\bm{l}}$ texture at the center of the vortex becomes more uniform and therefore vorticity-free due to the increasing energy cost of bending deformations. At high temperatures, the vortex center similarly becomes uniform, but this time as a result of the dipole interaction preferring the orientation $\hat{\bm{l}}\parallel\hat{\bm{d}} = \hat{\bm{z}}$, due to the transverse magnetic field.

As mentioned earlier, the axisymmetric ATC vortex is not the structure typically observed in realistic systems. However, the simple model calculations indicate that the logarithmic divergence of $K_b$ could cause textural changes in more complicated vortex systems as well. The specific changes will depend on the structure in question, but the formation of topological twist solitons seems like a good candidate for reducing bending energy, if possible. 

\section{\label{sec:vortexsheet}Vortex Sheet}

The first realistic structure we consider is the vortex sheet. The different merons inside a sheet are easily distinguishable, and aside from the asymptotic behaviour the circular merons bear some similarities with the ATC vortex, which indicates that the bending energy could be reduced by a transformation like the one observed in the model vortex.

The vortex sheet structure is constructed using vortex formation at a flow instability. The initial state at zero velocity is the so called  PanAm-texture, an in-plane distribution of $\hat{\bm l}$, where on one half of the sample circumference $\hat{\bm l}$ is directed inwards and on the other half outwards. In simulations, reorientation of $\hat{\bm l}$ happens within a disretization triangle and the full texture includes two such defects. In real $^3$He-A, these defects have a hard core and thus cannot be adequately represented in our London-limit calculations. Nevertheless, on increase of the rotation velocity in simulations we observe that the defects act as a source of vorticity, as in the experiments \cite{criticalvelocity}, and this is sufficient for our purposes. The radius of the calculation domain is $R = 115\,\mu$m, the magnetic field is applied along the $y$ axis and the temperature is $0.80T_{\rm c}$. The angular velocity $\Omega$ is increased gradually in steps of $0.1$ rad/s, using the minimization result of the previous step as the initial state of the next one. After two DQVs enter the system, they merge into a circular sheet with four quanta of circulation like the one in figure \ref{fig:vortexstructures}b. Then $\Omega$ is decreased to find the rotation velocity where the total energy is lowest, which occurs at $\Omega = 5.7$ rad/s. A detailed report on vortex formation and merging to sheets in our calculations will be published separately.

\begin{figure*}
    \centering
    \includegraphics[width=\linewidth,keepaspectratio]{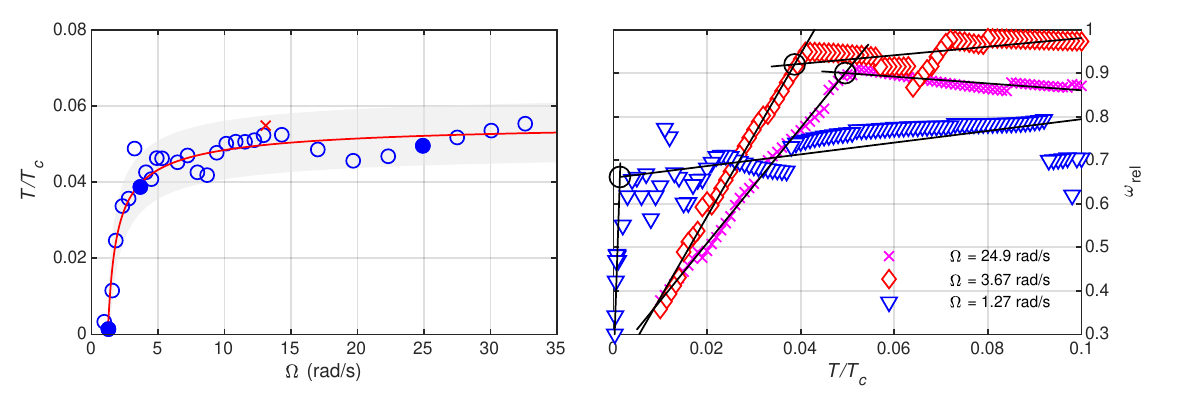}
    \caption{(\textit{Left}) An $\Omega$-$T$ phase diagram of the transition in the vortex sheet. The blue circles mark the transition temperatures at different values of $\Omega$. The solid red line shows a fit of the expression $A\exp(-B/(\Omega - \Omega_0)^{2/3})$ to the data points, with the $A = 0.056$, $B = 0.657\,(\text{rad/s})^{2/3}$ and $\Omega_0 = 1.15\,\text{rad/s}$. The red cross is the transition point calculated for a vortex sheet with 12 quanta of circulation, and the filled circles indicate the points from the example fits on the right. (\textit{Right}) Three example plots of relative vorticity $\omega_{\rm rel}$ as a function of temperature, taken from sweeps with $\Omega = 1.27$, $3.67$ and $24.9$ rad/s. The calculated points are marked with triangles, diamonds and crosses, respectively, and the solid lines on each sweep show the two straight lines that best fit the data. The transition point is taken as the intersection of these fits and is marked with a circle in each sweep. The corresponding transition temperatures are approximately $0.0014T_{\rm c}$, $0.039T_{\rm c}$ and $0.050T_{\rm c}$. The data from sweeps with lower $\Omega$ is more noisy, due to the reduced mesh resolution in the correspondingly larger cylinders.}
    \label{fig:phasediagram}
\end{figure*}

With a stable four-quanta sheet, the temperature is decreased down to $0.006T_{\rm c}$. During the temperature sweep, the sheet reconnects with the container walls, splitting into two separate sheets (see figure \ref{fig:vortexstructures}c), each with two quanta of circulation embedded in a splay soliton. In order to test the stability of the newly formed two-sheet texture, further temperature sweeps were performed starting from the lowest temperature of $0.006T_{\rm c}$ back up to $0.80T_{\rm c}$ and then down again. The two-sheet state persists.

A transition similar to the one discussed in section \ref{sec:modelvortex} occurs at temperatures below $0.05T_{\rm c}$. The textural transition is shown in figure \ref{fig:zerochargetransition}. At high temperatures (figure \ref{fig:zerochargetransition}a), the vortex sheet has the familiar structure with uniform vorticity along the sheet. As the temperature decreases, the vorticity becomes more concentrated at the merons (figure \ref{fig:zerochargetransition}d). The vorticity forms "bridges" between the circular and hyperbolic merons of the neighboring sheet, although the sheets still remain distinctly separate, as indicated by the $|\hat{\bm{l}}\times\hat{\bm{d}}|$ density plots in figure \ref{fig:zerochargetransition}c, f and i.

Further decreasing the temperature to below $0.05T_{\rm c}$, the $\hat{\bm{l}}$ texture of the circular merons, marked with the dashed lines in figure \ref{fig:zerochargetransition}, becomes more similar to the topological twist soliton. The meron center with vertical $\hat{\bm{l}}$ orientation increases in size, concentrating the bending deformation (and vorticity) into a tube. Comparing figures \ref{fig:modeltexture}d and \ref{fig:modeltexture}c to figures \ref{fig:zerochargetransition}e and \ref{fig:zerochargetransition}h shows the similarities between the two transitions. The transition is smooth and shows no hysteresis on repeated temperature sweeps.

\begin{figure*}
    \centering
    \includegraphics[width=\linewidth,keepaspectratio]{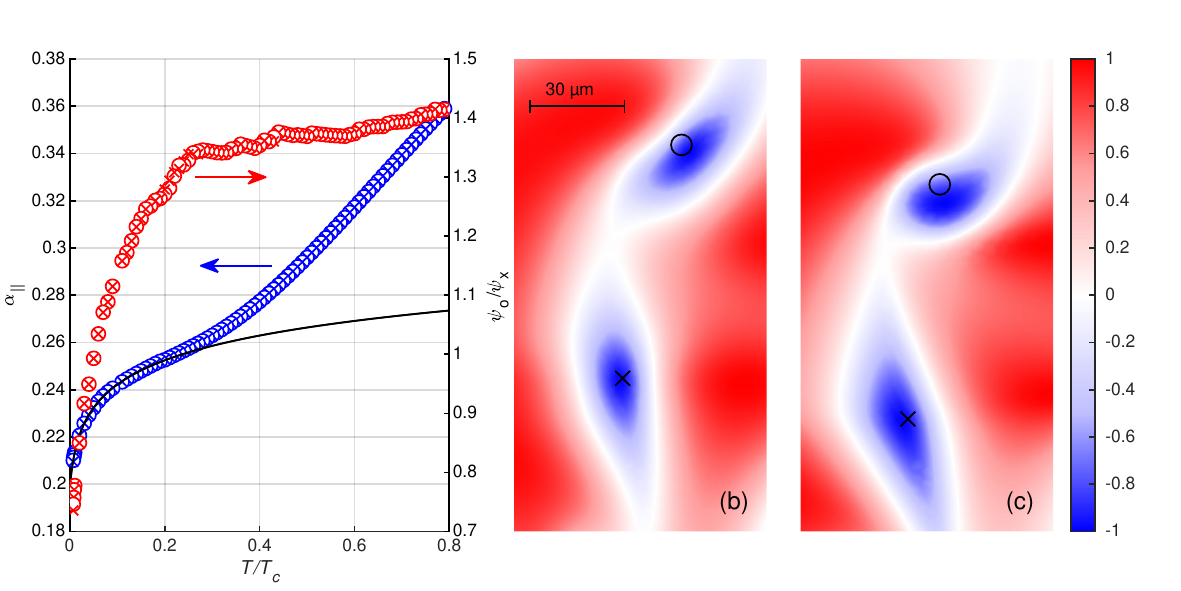}
    \caption{\label{fig:nmrplot} \textbf{(a)} The eigenvalue $\alpha_\parallel$ of the largest satellite peak in the calculated NMR spectrum as a function of temperature plotted as blue circles and crosses. The circles correspond to the upward temperature sweep started from $0.006T_{\rm c}$ while the crosses correspond to the return sweep back down from $0.80T_{\rm c}$. The near perfect match of the values of both sweeps indicates that there is no hysteresis in the transition. The solid black line is a logarithmic fit of the expression $A + B\ln(T/T_{\rm c} + C)$ to the eigenvalues below $0.20T_{\rm c}$, with $A = 0.277$, $B = 0.015$ and $C  = 0.007$. The red circles and crosses plotted against the right axis mark the ratio $|\psi_o|/|\psi_x|$ of the magnitude of the eigenfunctions at the circular and hyperbolic meron centers. \textbf{(b)} The potential \eqref{eq:nmrpotential} for spin waves produced by the vortex sheet on the left side of the cylinder in figure \ref{fig:zerochargetransition}d, at $0.20T_{\rm c}$. The circular and hyperbolic merons are difficult to distinguish visually. The centers of the circular and hyperbolic merons are marked with circles and crosses, respectively. \textbf{(c)} The potential for the vortex sheet in figure \ref{fig:zerochargetransition}g, at $0.006T_{\rm c}$. There is clear asymmetry between the shapes of spin-wave traps at the two merons.}
\end{figure*}

A change can be observed in the vorticity near the hyperbolic merons as well. The vorticity at low temperatures resembles a cross, with one line along the vortex sheet and one perpendicular to it. No quantitative analysis has been performed on the hyperbolic meron structure, but we believe that the change can be explained using the same reasoning as for the circular merons. The bending deformation in the hyperbolic meron is limited to the directions along these vorticity lines, while diagonal to the "arms" of the vorticity cross the $\hat{\bm{l}}$ vectors twist. In the hyperbolic merons the vorticity can be thought of as "thin-sheet" vorticity instead of tube vorticity.

The phase diagram in figure \ref{fig:phasediagram} is constructed by performing a temperature sweep for each rotation velocity. Initial states at different rotation velocities are found at $T=0.01T_{\rm c}$ by changing $\Omega$ while simultaneously adjusting the radius of the cylinder accordingly to avoid vortices entering or escaping. $\Omega$ is then a measure of the vortex density of the system. The radius is changed by interpolating the texture to a cylinder with different size.

Like for the model vortex, the relative vorticity $\omega_{\rm rel}$ is found to be a good quantitative indicator for the transition. Above the transition temperature the maximum vorticity in the system is found at the centers of the merons and $\omega_{\rm rel} \approx 1$. Below the transition temperature, $\omega_{\rm rel}$ decreases linearly as the $\hat{\bm{l}}$ texture becomes more uniform and the vorticity $\bm{\omega}$ at the center of the circular meron decreases according to the Mermin-Ho relation \eqref{eq:merminhorelation}. The linear decrease can be seen in the right side plot in figure \ref{fig:phasediagram}.

The transition temperature was found to be dependent on the vortex density of the system. A fit of the expression $A\exp(-B/(\Omega-\Omega_0)^{2/3})$ to the data gives the asymptotic transition temperature at high rotation velocities $T = 0.056 T_{\rm c}$. (The origin of the exponents 2/3 is discussed below.) At low velocities the transition temperature decreases and the zero temperature $\Omega$ cutoff is $\Omega_0 = 1.15\,\text{rad/s}$.

The calculation of the NMR response of the sheet at $\Omega = 5.7$ rad/s as a function of temperature is shown in figure \ref{fig:nmrplot}a. The eigenvalue $\alpha_\parallel$ of the most intense satellite peak in the NMR spectrum decreases linearly with temperature down to around $T=0.3T_{\rm c}$, after which the decrease becomes logarithmic $\alpha_{\parallel}\propto \ln(T/T_{\rm c})$. This indicates that the effects of the logarithmic divergence of the $K_b$ coefficient could be observable through NMR experiments, although the required temperatures are very low.

In the course of the transition the NMR potential distribution \eqref{eq:nmrpotential} changes. At temperatures above the transition the potential wells formed by the circular and hyperbolic merons in the sheet look very similar, shown in figure \ref{fig:nmrplot}b, while below the transition temperature there is a clear visual difference between the two (figure \ref{fig:nmrplot}c) with a larger potential well at the hyperbolic meron. Correspondingly during the transition the eigenfunction becomes more concentrated at the hyperbolic meron. The ratio between the magnitudes of the eigenfunction at the circular center $|\psi_o|$ and at the hyperbolic center $|\psi_x|$ is plotted in figure \ref{fig:nmrplot}a. At higher temperatures the ratio $|\psi_o|/|\psi_x| > 1$ , while $|\psi_o|/|\psi_x| < 1$ at low temperatures. Notably the transition temperature seems to correspond roughly to the point where the ratio is close to one. This could be used as another indicator for the transition, although it is more indirect than the one used above.

The transition to the tube vorticity state for the vortex sheet can be explained qualitatively using similar reasoning that was used for the ATC vortex: the bending energy is reduced by confining the deformations to a narrow tube around the center of the circular meron with uniformly oriented $\hat{\bm{l}}$. In the vortex sheet the largest relevant length scale for the $\hat{\bm{l}}$ vector gradients is the intermeron distance $p$ along the sheet. The $\Omega^{-2/3}$ dependence of the transition temperature in the phase diagram of figure \ref{fig:phasediagram} could be explained by the fact that $p \propto \Omega^{-1/3}$ and above the transition the elastic energy density is proportional to $p^{-2} \propto \Omega^{2/3}$. As the vortex density increases, the bending energy contribution from intermeron gradients becomes larger and can be reduced by the formation of vorticity tubes even at higher temperatures and smaller values of $K_b$. According to the fit in the phase diagram in figure \ref{fig:phasediagram}, at $T = 0$ the transition occurs at a finite vortex density corresponding to $\Omega = 1.15\,\text{rad/s}$.

\begin{figure}[b]
    \centering
    \includegraphics[width=\linewidth,keepaspectratio]{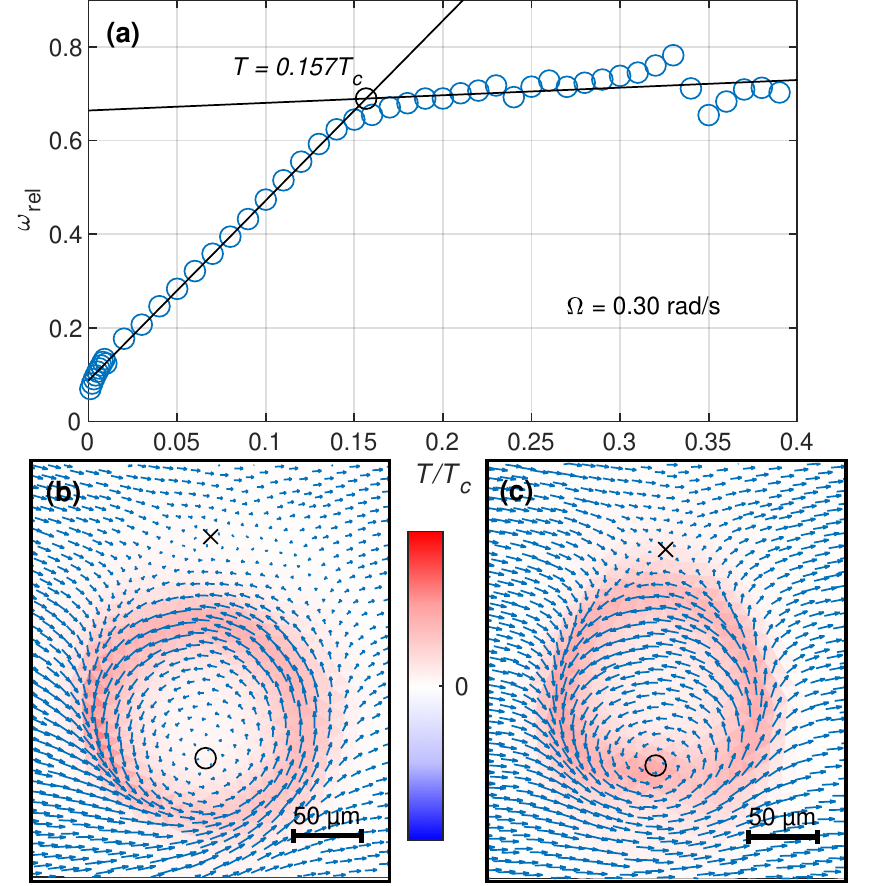}
    \caption{\label{fig:vortexzerochargetransition} The transition in separate double-quantum vortices at $\Omega = 0.30$ rad/s. \textbf{(a)} Plot of $\omega_{\rm rel}$ as a function of temperature. The solid lines are the best linear fits to the data. The transition point $T = 0.157T_{\rm c}$ is taken as the intersection of these lines. \textbf{(b)} The $\hat{\bm{l}}$-vector texture of the double-quantum vortex below the transition temperature at $0.001T_{\rm c}$. The color indicates vorticity $\bm{\omega}$. The centers of the circular and hyperbolic meron are marked with a circle and a cross, respectively. \textbf{(c)} The texture at $0.20T_{\rm c}$. }
\end{figure}

\section{\label{sec:doublequantumvortices}Double-quantum vortices}

\begin{figure*}
    \centering
    \includegraphics[width=\linewidth,keepaspectratio]{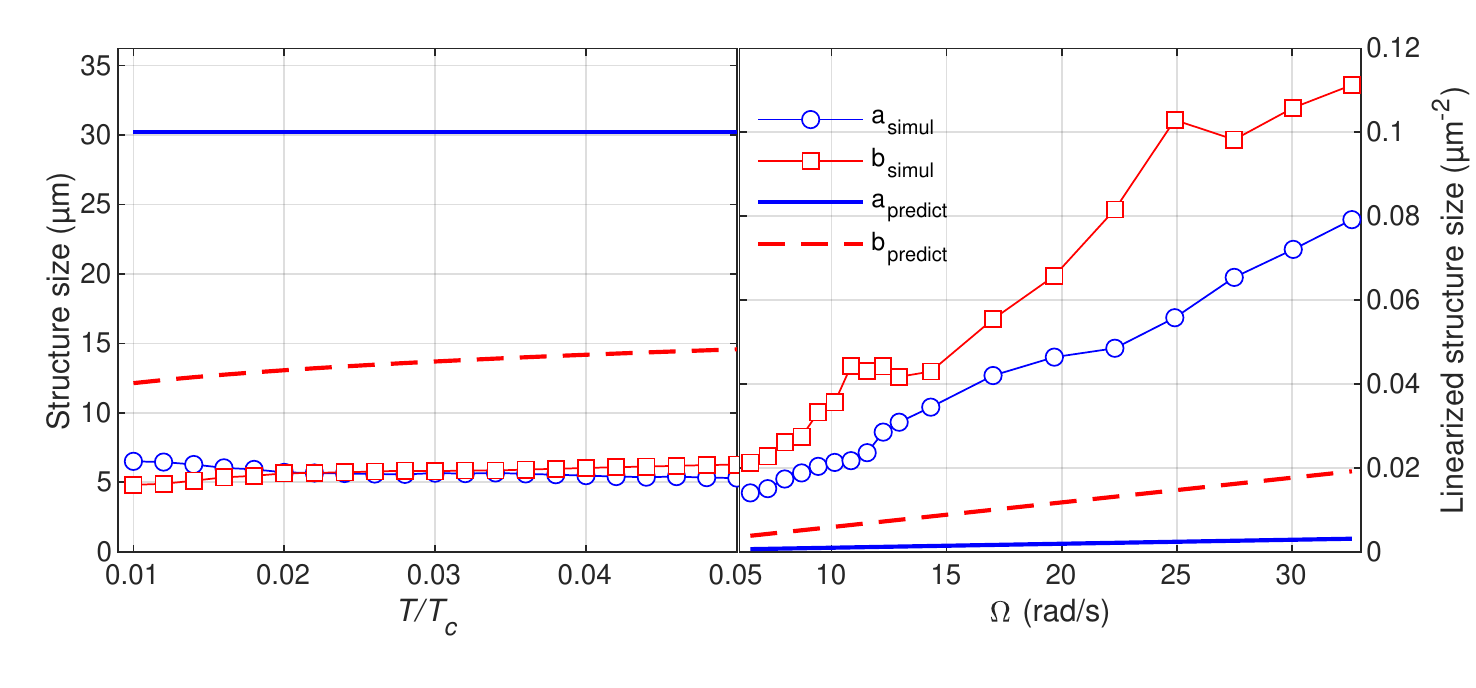}
    \caption{\label{fig:sheetparameters} The radius $a$ and width $b$ of the circular meron vorticity tubes in the vortex sheet. The solid line and the dashed line are the predicted values of $a$ and $b$, respectively, while the circles and squares correspond to their numerically calculated values. (\textit{Left}) $a$ and $b$ as a function of temperature, taken from the sweep done at $\Omega = 11.55$ rad/s. The value of $\Omega$ was chosen such that a wide enough temperature range was available below the transition temperature. The predicted values are much higher than calculated, almost six times higher for $a$ and three times for $b$. The temperature dependence of $b$ agrees with the prediction, while $a$ decreases slightly with increasing temperature. (\textit{Right}) $a^{-2}$ and $b^{-2}$ as functions of $\Omega$ at $T = 0.01T_{\rm c}$. The calculated values are approximately linear in these coordinates, as predicted.}
\end{figure*}

The non-axisymmetric double-quantum vortex (see figure \ref{fig:vortexstructures}a) is the most common topological object formed in $^3$He-A\cite{DQVNature}. The vorticity in a DQV is distributed in a tube around the vortex axis at all temperatures, so a priori it is difficult to determine what the qualitative effect the logarithmic divergence of the $K_b$ coefficient would have on its structure. However, in the region between the two merons the texture is similar to the ATC vortex. On the line between the hyperbolic and circular meron centers, $\hat{\bm{l}}$ rotates with a twist-type deformation over a distance $d$, while across the vortex (perpendicular to the line between merons) the deformation is splay/bend type over a distance $w$. The elastic energy is then $F_{\rm el}\sim K_b(d/w) + K_t(w/d)$ and at low temperatures where $K_b \gg K_t$, the energy is minimized by decreasing the width $d$ of the region between $\hat{\bm{l}} = \hat{\bm{z}}$ and $\hat{\bm{l}} = -\hat{\bm{z}}$. In the whole DQV texture this is seen as a shift in the location of the vorticity tube, so that it is centered around the circular meron instead of the vortex axis. Then the size of the region around the meron center where $\hat{\bm{l}}$ is oriented along the vertical axis is again expected to increase, while the twist rotation occurs in a thin region between the two merons.

A low-temperature state with two separated double-quantum vortices has been created as follows: a clean two-vortex state is found from an $\Omega$ sweep done in the axial field, starting with a PanAm-texture in a larger cylinder with radius $R=500\,\mu$m at $0.80T_{\rm c}$. The larger cylinder is chosen to keep applied rotation velocity below the treshold for merging vortices to sheets, and to accomodate the lower $\Omega$ values, the $\Omega$ sweep is done in smaller increments of $0.01$ rad/s. When the first vortices have entered and found stable locations near the center of the cylinder at $\Omega=1.27$ rad/s, the temperature is reduced down to $0.001T_{\rm c}$, the magnetic field rotated to the transverse direction and $\Omega$ reduced to $0.30$ rad/s where the energy is at a minimum.

A temperature sweep is performed in increasing direction on this state with two double-quantum vortices. At the start of the sweep at $0.001T_{\rm c}$, the vorticity $\bm{\omega}$ in the vortices is concentrated into a tube shape around the circular meron, with barely any vorticity near the hyperbolic meron, as shown in figure~\ref{fig:vortexzerochargetransition}b. The value of $\omega_{\rm rel}$ calculated at the center of the circular meron at these temperatures is close to zero.

On increase of temperature, the value of $\omega_{\rm rel}$ increases linearly at low temperatures and stays constant above $0.15T_{\rm c}$, as shown in figure \ref{fig:vortexzerochargetransition}a. Thus we find the transition of the same type as for the ATC vortex and the vortex sheet. However, the transition temperature for vortices is different from that for vortex sheets in the phase diagram in figure \ref{fig:phasediagram} at the same velocity. At such low velocities, the vortex sheet doesn't appear to have a transition at all.

Above the transition temperature the vortices look like well-known $w$-vortices with a hyperbolic and circular meron (figure \ref{fig:vortexzerochargetransition}c). The vorticity is spread in a tube shape even at high temperatures, but the tubes are centered around the axis of the whole two quanta structure. Below the transition point the tube shifts and becomes centered around only the circular meron (figure \ref{fig:vortexzerochargetransition}b).

Qualitatively the low temperature texture resembles the ATC vortex: the core region of the circular meron is highly uniform, in order to minimize the region where $\hat{\bm{l}}$ bends. In a finite-radius cylinder, however, the $\hat{\bm{l}}$ vectors far from the vortex are horizontal instead of vertical, due to the orienting effect of the container walls.

\section{\label{sec:modelcomparison}Comparison with the ATC vortex}

\begin{figure}
    \centering
    \includegraphics[width=\linewidth,keepaspectratio]{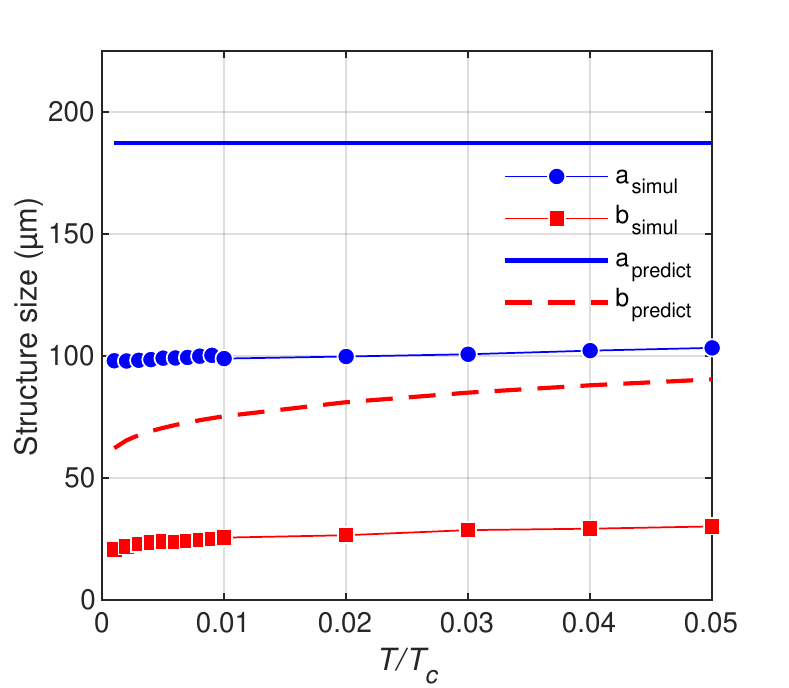}
    \caption{\label{fig:vortexparameters} The radius $a$ and width $b$ of the vorticity tubes in separated double-quantum vortices as a function of temperature. The predicted $a$ and $b$ are marked by a solid and dashed line, respectively. The measured values are marked with circles and squares for $a$ and $b$, respectively. The predicted values are higher by approximately a factor of 2. }
\end{figure}

The appearance of the tube vorticity distribution in the circular merons in vortex sheets and double-quantum vortices agrees qualitatively with the model ATC vortex in section \ref{sec:modelvortex}. However, the quantitative prediction for the size of the tubes does not match well, as shown in figure \ref{fig:sheetparameters} for the vortex sheet and in figure \ref{fig:vortexparameters} for separated vortices. For sheets, the value of $a$ is almost six times lower than predicted, while $b$ is almost three times lower. In vortices the numerically calculated values differ by approximately a factor of two from the predicted values.

The lower measured values can be at least partially explained qualitatively. The full simulations include the $C$-terms omitted in the model derivation, which were found to be highly impactful in section \ref{sec:modelvortex}. Additionally, the ATC vortex structure in the model assumed an axisymmetric structure with the bulk $\hat{\bm{l}}$ texture being uniformly vertical. In realistic situations, the finite size of the domain restricts the bulk texture to be in-plane due to the effects of the boundary conditions. In this case the bending energy density cannot be strictly concentrated into a narrow tube, because outside the meron core there will be some splay/bend distortion in the bulk texture. Finally, the repulsive effect of neighboring quanta of circulation is expected to reduce the size of the tube by a factor that is dependent on the distance between quanta.

In the vortex sheet, the tubes form around the circular merons, which are single circulation quantum structures instead of the $\nu = 2$ ATC vortex considered in the model. The adjustment in the model equations \eqref{eq:a} for $a$ and \eqref{eq:b} for $b$ is done naively by assuming a superfluid velocity outside the vortex is twice smaller than in the original derivation. As mentioned previously, the change in number of quanta has an additional indirect effect on the calculated values through the change in the asymptotic behaviour of $\hat{\bm{l}}$ vectors outside the vortex (horizontal vs. vertical). 

\section{\label{sec:conclusion}Conclusion}

We have numerically calculated equlibrium order-parameter textures in rotating $^3$He-A at low temperatures where the effect of the logarithmic divergence of the bending coefficient $K_b$ in the free energy is relevant. The connection of this divergence to the zero-charge effect of quantum electrodynamics and the appearance of vorticity tubes at low temperatures was predicted by Volovik \cite{volovik}. A transition to the predicted state has been found both in the vortex sheet and in separate vortices. The transition temperature is found to depend on the vortex density of the system, and a temperature-vortex density phase diagram has been presented for the vortex sheet.

In our calculations in the absense of pinning, vortices are stable only with applied rotation. The original prediction of Ref.~\cite{volovik} has been adjusted to include the effect of rotation. The analytic model, nevertheless, does not capture all the details of the realistic textures and the size of the vorticity tubes in the simulated textures is considerably smaller than that in the model. In particular, the so-called $C$ term in the superfluid velocity, ignored in the model, turned out to play an important role in shaping vortex structures. Another important difference between the model and the realistic textures is the asymptotic behaviour of $\hat{\bm{l}}$ at large radii, where the model ignores solid-wall boundary conditions. The calculations also have their limitations: They are done with the assumption of a uniform texture in the $z$ direction, which means that possible three dimensional structures, related to the axial superflow in broken-symmetry vortex cores, could not be found.

In search of observable signatures of the transition, we have calculated 
the NMR response of the vortex sheet as a function of temperature. As expected, restructuring of the distribution of vorticity has a profound effect on the frequency shift of the characteristic satellite in the NMR spectrum: The satellite moves further from the bulk peak towards the Larmor value. The logarithmic dependence of the frequency shift, reflecting that of $K_b$ becomes prominent only at temperatures below $0.2\,T_{\rm c}$ which may make observation of this effect in experiments challenging.

\begin{acknowledgments}
We thank Grigory Volovik, Erkki Thuneberg and Jaakko Nissinen for stimulating discussions. This work has been supported by the European Research Council (ERC) under the European Union's Horizon 2020 research and innovation programme (Grant Agreement No. 694248) and by Academy of Finland (grant 332964).

We acknowledge the computational resources provided by the Aalto Science-IT project.
\end{acknowledgments}

\appendix

\section{\label{sec:appendix}Parameterization of the order parameter}
Quaternions are an extension of the complex number system into four dimensions and are of the form
\begin{equation}
    q = q_0 + q_1i + q_2j + q_3k
    \label{eq:quaternionform}
\end{equation}
with imaginary units $i$, $j$ and $k$ defined by the relation
\begin{equation}
    i^2 = j^2 = k^2 = ijk = -1.
    \label{eq:quaternionunits}
\end{equation}
Sometimes it is useful to use the notation
\begin{equation}
    q = q_0 + \bm{q}
    \label{eq:quaternionvectorform}
\end{equation}
where $q_0$ is called the real part and $\bm{q}$ the vector part.

Three dimensional rotations and orientations can be described by quaternions, analoguously to how complex numbers can be used to represent two dimensional rotations. A rotation in 3D defined by an unit vector axis $\bm{u}$ and an angle $\theta$ can be expressed as a unit quaternion
\begin{equation}
    q = \cos\frac{\theta}{2} + \sin\frac{\theta}{2}\bm{u}
    \label{eq:quaternionrotation}
\end{equation}

The orientation of the orthonormal orbital triad $(\hat{\bm{m}}, \hat{\bm{n}}, \hat{\bm{l}})$ can be represented with a single quaternion using the conversion formula for rotation matrices:
\begin{align}
    &\begin{bmatrix}
    m_x & n_x & l_x \\
    m_y & n_y & l_y \\
    m_z & n_z & l_z
    \end{bmatrix} \nonumber \\ &= 
    \begin{bmatrix}
    1-2q_2^2-2q_3^2 & 2(q_1q_2-q_3q_0) & 2(q_1q_3+q_2q_0) \\
    2(q_1q_2+q_3q_0) & 1-2q_1^2-2q_3^2 & 2(q_2q_3-q_1q_0) \\
    2(q_1q_3-q_2q_0) & 2(q_1q_0+q_2q_3) & 1-2q_1^2-2q_2^2
    \end{bmatrix}
    \label{eq:quattorot}
\end{align}

The benefit of quaternions over other rotation formalisms is that they reduce the number of required parameters from nine to four, and they can describe any orientation without singularities or gimbal lock.

The spin anisotropy vector $\hat{\bm{d}}$ is parameterized with azimuthal and polar angles $\alpha$ and $\beta$. To avoid issues when $\beta = 0$, the polar axis is chosen as the magnetic field direction $\bm{H}$. In our system the $\bm{H}$ vector is confined to the $yz$ plane and its direction is described by an angle $\mu$ between $\bm{H}$ and the $z$-axis. The $\hat{\bm{d}}$ vector can then be parameterized as
\begin{align}
    d_x &= \cos{\alpha}\sin{\beta} \nonumber \\
    d_y &= \cos{\beta}\sin{\mu}-\cos{\mu}\sin{\alpha}\sin{\beta} \\
    d_z &= \cos{\beta}\cos{\mu}+\sin{\alpha}\sin{\beta}\sin{\mu} \nonumber
    \label{eq:dvector}
\end{align}
The magnetic field direction is kept static during minimization, so $\mu$ is a constant.

The quaternion and $\hat{\bm{d}}$ angle values are defined at each node in the mesh. To calculate the energy for a single triangle, the parameters are linearly interpolated to quadrature points using barycentric coordinates, where the energy densities are computed. The integration is then performed using Gaussian quadrature rules. After the quaternions are interpolated, they must be renormalized to keep them unit length. 

\section{\label{sec:coefficients}Coefficients}
\begin{figure}[b]
\includegraphics[width=\linewidth,keepaspectratio]{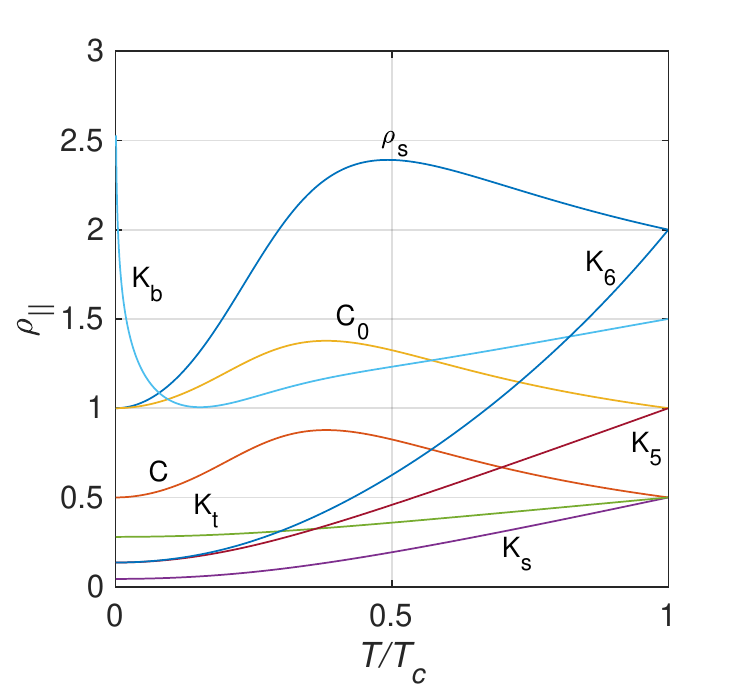}
\caption{\label{fig:coefficients} The coefficients values used in the energy calculation as a function of temperature. The coefficients are normalized to $\rho_\parallel$ to better visualize their relationships.}
\end{figure}

The coefficients for the energy densities in Equations \eqref{eq:fdip}, \eqref{eq:fmag}, \eqref{eq:fkin} and \eqref{eq:fel} are presented in Figure \ref{fig:coefficients}. The values are normalized to $\rho_\parallel$ in order to demonstrate their relative behaviour. Note the logarithmic divergence of the bending coefficient $K_b$ as $T\rightarrow 0$.

\bibliography{zerocharge}

\begin{thebibliography}{38}%
\makeatletter
\providecommand \@ifxundefined [1]{%
 \@ifx{#1\undefined}
}%
\providecommand \@ifnum [1]{%
 \ifnum #1\expandafter \@firstoftwo
 \else \expandafter \@secondoftwo
 \fi
}%
\providecommand \@ifx [1]{%
 \ifx #1\expandafter \@firstoftwo
 \else \expandafter \@secondoftwo
 \fi
}%
\providecommand \natexlab [1]{#1}%
\providecommand \enquote  [1]{``#1''}%
\providecommand \bibnamefont  [1]{#1}%
\providecommand \bibfnamefont [1]{#1}%
\providecommand \citenamefont [1]{#1}%
\providecommand \href@noop [0]{\@secondoftwo}%
\providecommand \href [0]{\begingroup \@sanitize@url \@href}%
\providecommand \@href[1]{\@@startlink{#1}\@@href}%
\providecommand \@@href[1]{\endgroup#1\@@endlink}%
\providecommand \@sanitize@url [0]{\catcode `\\12\catcode `\$12\catcode
  `\&12\catcode `\#12\catcode `\^12\catcode `\_12\catcode `\%12\relax}%
\providecommand \@@startlink[1]{}%
\providecommand \@@endlink[0]{}%
\providecommand \url  [0]{\begingroup\@sanitize@url \@url }%
\providecommand \@url [1]{\endgroup\@href {#1}{\urlprefix }}%
\providecommand \urlprefix  [0]{URL }%
\providecommand \Eprint [0]{\href }%
\providecommand \doibase [0]{https://doi.org/}%
\providecommand \selectlanguage [0]{\@gobble}%
\providecommand \bibinfo  [0]{\@secondoftwo}%
\providecommand \bibfield  [0]{\@secondoftwo}%
\providecommand \translation [1]{[#1]}%
\providecommand \BibitemOpen [0]{}%
\providecommand \bibitemStop [0]{}%
\providecommand \bibitemNoStop [0]{.\EOS\space}%
\providecommand \EOS [0]{\spacefactor3000\relax}%
\providecommand \BibitemShut  [1]{\csname bibitem#1\endcsname}%
\let\auto@bib@innerbib\@empty
\bibitem [{\citenamefont {Vollhardt}\ and\ \citenamefont
  {Wölfle}(1990)}]{VollhardtWolfle}%
  \BibitemOpen
  \bibfield  {author} {\bibinfo {author} {\bibfnamefont {D.}~\bibnamefont
  {Vollhardt}}\ and\ \bibinfo {author} {\bibfnamefont {P.}~\bibnamefont
  {Wölfle}},\ }\href {https://doi.org/10.1201/b12808} {\emph {\bibinfo {title}
  {The Superfluid Phases of Helium 3}}}\ (\bibinfo  {publisher}
  {Taylor\&Francis},\ \bibinfo {address} {London},\ \bibinfo {year}
  {1990})\BibitemShut {NoStop}%
\bibitem [{\citenamefont {Leggett}(1975)}]{LeggettReview}%
  \BibitemOpen
  \bibfield  {author} {\bibinfo {author} {\bibfnamefont {A.~J.}\ \bibnamefont
  {Leggett}},\ }\bibfield  {title} {\bibinfo {title} {A theoretical description
  of the new phases of liquid $^{3}\mathrm{He}$},\ }\href
  {https://doi.org/10.1103/RevModPhys.47.331} {\bibfield  {journal} {\bibinfo
  {journal} {Rev. Mod. Phys.}\ }\textbf {\bibinfo {volume} {47}},\ \bibinfo
  {pages} {331} (\bibinfo {year} {1975})}\BibitemShut {NoStop}%
\bibitem [{\citenamefont {Walmsley}\ and\ \citenamefont
  {Golov}(2012)}]{chiralsuperfluid}%
  \BibitemOpen
  \bibfield  {author} {\bibinfo {author} {\bibfnamefont {P.~M.}\ \bibnamefont
  {Walmsley}}\ and\ \bibinfo {author} {\bibfnamefont {A.~I.}\ \bibnamefont
  {Golov}},\ }\bibfield  {title} {\bibinfo {title} {Chirality of superfluid
  $^{3}\mathrm{He}$-$\mathrm{A}$},\ }\href
  {https://doi.org/10.1103/PhysRevLett.109.215301} {\bibfield  {journal}
  {\bibinfo  {journal} {Phys. Rev. Lett.}\ }\textbf {\bibinfo {volume} {109}},\
  \bibinfo {pages} {215301} (\bibinfo {year} {2012})}\BibitemShut {NoStop}%
\bibitem [{\citenamefont {Volovik}(2010)}]{VolovikBook}%
  \BibitemOpen
  \bibfield  {author} {\bibinfo {author} {\bibfnamefont {G.}~\bibnamefont
  {Volovik}},\ }\href
  {https://doi.org/10.1093/acprof:oso/9780199564842.001.0001} {\emph {\bibinfo
  {title} {The Universe in a Helium Droplet}}}\ (\bibinfo  {publisher} {Oxford
  University Press},\ \bibinfo {address} {United Kingdom},\ \bibinfo {year}
  {2010})\BibitemShut {NoStop}%
\bibitem [{\citenamefont {Bevan}\ \emph {et~al.}(1997)\citenamefont {Bevan},
  \citenamefont {Manninen}, \citenamefont {Cook}, \citenamefont {Hook},
  \citenamefont {Hall}, \citenamefont {Vachaspati},\ and\ \citenamefont
  {Volovik}}]{bevannature}%
  \BibitemOpen
  \bibfield  {author} {\bibinfo {author} {\bibfnamefont {T.}~\bibnamefont
  {Bevan}}, \bibinfo {author} {\bibfnamefont {A.}~\bibnamefont {Manninen}},
  \bibinfo {author} {\bibfnamefont {J.}~\bibnamefont {Cook}}, \bibinfo {author}
  {\bibfnamefont {J.}~\bibnamefont {Hook}}, \bibinfo {author} {\bibfnamefont
  {H.}~\bibnamefont {Hall}}, \bibinfo {author} {\bibfnamefont {T.}~\bibnamefont
  {Vachaspati}},\ and\ \bibinfo {author} {\bibfnamefont {G.}~\bibnamefont
  {Volovik}},\ }\bibfield  {title} {\bibinfo {title} {Momentum creation by
  vortices in superfluid $^3\mathrm{He}$ as a model of primodial
  baryogenesis},\ }\href {https://doi.org/10.1038/386689a0} {\bibfield
  {journal} {\bibinfo  {journal} {Nature}\ }\textbf {\bibinfo {volume} {386}},\
  \bibinfo {pages} {689} (\bibinfo {year} {1997})}\BibitemShut {NoStop}%
\bibitem [{\citenamefont {Balatskii}\ \emph {et~al.}(1986)\citenamefont
  {Balatskii}, \citenamefont {Volovik},\ and\ \citenamefont
  {Konyshev}}]{chiralanomaly}%
  \BibitemOpen
  \bibfield  {author} {\bibinfo {author} {\bibfnamefont {A.~V.}\ \bibnamefont
  {Balatskii}}, \bibinfo {author} {\bibfnamefont {G.~E.}\ \bibnamefont
  {Volovik}},\ and\ \bibinfo {author} {\bibfnamefont {V.~A.}\ \bibnamefont
  {Konyshev}},\ }\bibfield  {title} {\bibinfo {title} {On the chiral anomaly in
  superfluid $^3\mathrm{He}$-$\mathrm{A}$},\ }\href@noop {} {\bibfield
  {journal} {\bibinfo  {journal} {Sov. Phys. JETP}\ }\textbf {\bibinfo {volume}
  {63}},\ \bibinfo {pages} {1194} (\bibinfo {year} {1986})}\BibitemShut
  {NoStop}%
\bibitem [{\citenamefont {Ishihara}\ \emph {et~al.}(2019)\citenamefont
  {Ishihara}, \citenamefont {Mizushima}, \citenamefont {Tsuruta},\ and\
  \citenamefont {Fujimoto}}]{TCME}%
  \BibitemOpen
  \bibfield  {author} {\bibinfo {author} {\bibfnamefont {Y.}~\bibnamefont
  {Ishihara}}, \bibinfo {author} {\bibfnamefont {T.}~\bibnamefont {Mizushima}},
  \bibinfo {author} {\bibfnamefont {A.}~\bibnamefont {Tsuruta}},\ and\ \bibinfo
  {author} {\bibfnamefont {S.}~\bibnamefont {Fujimoto}},\ }\bibfield  {title}
  {\bibinfo {title} {Torsional chiral magnetic effect due to skyrmion textures
  in a {Weyl} superfluid $^{3}\mathrm{He}\text{\ensuremath{-}}\mathrm{A}$},\
  }\href {https://doi.org/10.1103/PhysRevB.99.024513} {\bibfield  {journal}
  {\bibinfo  {journal} {Phys. Rev. B}\ }\textbf {\bibinfo {volume} {99}},\
  \bibinfo {pages} {024513} (\bibinfo {year} {2019})}\BibitemShut {NoStop}%
\bibitem [{\citenamefont {Nissinen}\ and\ \citenamefont
  {Volovik}(2020)}]{Nissinen2020}%
  \BibitemOpen
  \bibfield  {author} {\bibinfo {author} {\bibfnamefont {J.}~\bibnamefont
  {Nissinen}}\ and\ \bibinfo {author} {\bibfnamefont {G.~E.}\ \bibnamefont
  {Volovik}},\ }\bibfield  {title} {\bibinfo {title} {Thermal nieh-yan anomaly
  in weyl superfluids},\ }\href
  {https://doi.org/10.1103/PhysRevResearch.2.033269} {\bibfield  {journal}
  {\bibinfo  {journal} {Phys. Rev. Research}\ }\textbf {\bibinfo {volume}
  {2}},\ \bibinfo {pages} {033269} (\bibinfo {year} {2020})}\BibitemShut
  {NoStop}%
\bibitem [{\citenamefont {Volovik}\ and\ \citenamefont
  {Mineev}(1981)}]{zeroTnormalcomponent1}%
  \BibitemOpen
  \bibfield  {author} {\bibinfo {author} {\bibfnamefont {G.~E.}\ \bibnamefont
  {Volovik}}\ and\ \bibinfo {author} {\bibfnamefont {V.~P.}\ \bibnamefont
  {Mineev}},\ }\bibfield  {title} {\bibinfo {title} {Orbital angular momentum
  and orbital dynamics: $^3\mathrm{He}$-$\mathrm{A}$ and the {Bose} liquid},\
  }\href@noop {} {\bibfield  {journal} {\bibinfo  {journal} {Sov. Phys. JETP}\
  }\textbf {\bibinfo {volume} {54}},\ \bibinfo {pages} {524} (\bibinfo {year}
  {1981})}\BibitemShut {NoStop}%
\bibitem [{\citenamefont {Dombre}\ and\ \citenamefont
  {Combescot}(1984)}]{zeroTnormalcomponent2}%
  \BibitemOpen
  \bibfield  {author} {\bibinfo {author} {\bibfnamefont {T.}~\bibnamefont
  {Dombre}}\ and\ \bibinfo {author} {\bibfnamefont {R.}~\bibnamefont
  {Combescot}},\ }\bibfield  {title} {\bibinfo {title} {Excitation spectrum and
  superfluid density of $^{3}\mathrm{He}$-$\mathrm{A}$ at $\mathit{T}=0$},\
  }\href {https://doi.org/10.1103/PhysRevB.30.3765} {\bibfield  {journal}
  {\bibinfo  {journal} {Phys. Rev. B}\ }\textbf {\bibinfo {volume} {30}},\
  \bibinfo {pages} {3765} (\bibinfo {year} {1984})}\BibitemShut {NoStop}%
\bibitem [{\citenamefont {Mermin}\ and\ \citenamefont
  {Muzikar}(1980)}]{anomalouscurrent}%
  \BibitemOpen
  \bibfield  {author} {\bibinfo {author} {\bibfnamefont {N.~D.}\ \bibnamefont
  {Mermin}}\ and\ \bibinfo {author} {\bibfnamefont {P.}~\bibnamefont
  {Muzikar}},\ }\bibfield  {title} {\bibinfo {title} {Cooper pairs versus
  {Bose} condensed molecules: The ground-state current in superfluid
  $^{3}\mathrm{He}$-$\mathrm{A}$},\ }\href
  {https://doi.org/10.1103/PhysRevB.21.980} {\bibfield  {journal} {\bibinfo
  {journal} {Phys. Rev. B}\ }\textbf {\bibinfo {volume} {21}},\ \bibinfo
  {pages} {980} (\bibinfo {year} {1980})}\BibitemShut {NoStop}%
\bibitem [{\citenamefont {Nissinen}\ and\ \citenamefont
  {Volovik}(2022)}]{Nissinen2022}%
  \BibitemOpen
  \bibfield  {author} {\bibinfo {author} {\bibfnamefont {J.}~\bibnamefont
  {Nissinen}}\ and\ \bibinfo {author} {\bibfnamefont {G.~E.}\ \bibnamefont
  {Volovik}},\ }\bibfield  {title} {\bibinfo {title} {Anomalous chiral
  transport with vorticity and torsion: Cancellation of two mixed gravitational
  anomaly currents in rotating chiral $p+ip$ weyl condensates},\ }\href
  {https://doi.org/10.1103/PhysRevD.106.045022} {\bibfield  {journal} {\bibinfo
   {journal} {Phys. Rev. D}\ }\textbf {\bibinfo {volume} {106}},\ \bibinfo
  {pages} {045022} (\bibinfo {year} {2022})}\BibitemShut {NoStop}%
\bibitem [{\citenamefont {Combescot}\ and\ \citenamefont
  {Dombre}(1983)}]{nonanalytic1}%
  \BibitemOpen
  \bibfield  {author} {\bibinfo {author} {\bibfnamefont {R.}~\bibnamefont
  {Combescot}}\ and\ \bibinfo {author} {\bibfnamefont {T.}~\bibnamefont
  {Dombre}},\ }\bibfield  {title} {\bibinfo {title} {Superfluid current in
  $^{3}\mathrm{He}$-$\mathrm{A}$ at $\mathit{T}=0$},\ }\href
  {https://doi.org/10.1103/PhysRevB.28.5140} {\bibfield  {journal} {\bibinfo
  {journal} {Phys. Rev. B}\ }\textbf {\bibinfo {volume} {28}},\ \bibinfo
  {pages} {5140} (\bibinfo {year} {1983})}\BibitemShut {NoStop}%
\bibitem [{\citenamefont {Cross}(1975)}]{nonanalytic2}%
  \BibitemOpen
  \bibfield  {author} {\bibinfo {author} {\bibfnamefont {M.~C.}\ \bibnamefont
  {Cross}},\ }\bibfield  {title} {\bibinfo {title} {A generalized
  {Ginzburg-Landau} approach to the superfluidity of helium 3},\ }\href
  {https://doi.org/10.1007/BF01141607} {\bibfield  {journal} {\bibinfo
  {journal} {J. Low Temp. Phys.}\ }\textbf {\bibinfo {volume} {21}},\ \bibinfo
  {pages} {525} (\bibinfo {year} {1975})}\BibitemShut {NoStop}%
\bibitem [{\citenamefont {Volovik}(1987)}]{zerocharge}%
  \BibitemOpen
  \bibfield  {author} {\bibinfo {author} {\bibfnamefont {G.~E.}\ \bibnamefont
  {Volovik}},\ }\bibfield  {title} {\bibinfo {title} {Peculiarities in the
  dynamics of superfluid $^3\mathrm{He}$-$\mathrm{A}$: analog of chiral anomaly
  and of zero-charge},\ }\href@noop {} {\bibfield  {journal} {\bibinfo
  {journal} {Sov. Phys. JETP}\ }\textbf {\bibinfo {volume} {65}},\ \bibinfo
  {pages} {1193} (\bibinfo {year} {1987})}\BibitemShut {NoStop}%
\bibitem [{\citenamefont {Hensley}\ \emph {et~al.}(1993)\citenamefont
  {Hensley}, \citenamefont {Lee}, \citenamefont {Hamot}, \citenamefont
  {Mizusaki},\ and\ \citenamefont {Halperin}}]{F0a}%
  \BibitemOpen
  \bibfield  {author} {\bibinfo {author} {\bibfnamefont {H.}~\bibnamefont
  {Hensley}}, \bibinfo {author} {\bibfnamefont {Y.}~\bibnamefont {Lee}},
  \bibinfo {author} {\bibfnamefont {P.}~\bibnamefont {Hamot}}, \bibinfo
  {author} {\bibfnamefont {T.}~\bibnamefont {Mizusaki}},\ and\ \bibinfo
  {author} {\bibfnamefont {W.}~\bibnamefont {Halperin}},\ }\bibfield  {title}
  {\bibinfo {title} {Measurement of the magnetic susceptibility of normal fluid
  $^3\mathrm{He}$ at very low temperatures},\ }\href
  {https://doi.org/10.1007/BF00682015} {\bibfield  {journal} {\bibinfo
  {journal} {J. Low Temp. Phys.}\ }\textbf {\bibinfo {volume} {90}},\ \bibinfo
  {pages} {149} (\bibinfo {year} {1993})}\BibitemShut {NoStop}%
\bibitem [{\citenamefont {Williams}\ and\ \citenamefont
  {Fetter}(1979)}]{slowlyrotating}%
  \BibitemOpen
  \bibfield  {author} {\bibinfo {author} {\bibfnamefont {M.~R.}\ \bibnamefont
  {Williams}}\ and\ \bibinfo {author} {\bibfnamefont {A.~L.}\ \bibnamefont
  {Fetter}},\ }\bibfield  {title} {\bibinfo {title} {Textures in slowly
  rotating $^{3}\mathrm{He}$-$\mathrm{A}$},\ }\href
  {https://doi.org/10.1103/PhysRevB.20.169} {\bibfield  {journal} {\bibinfo
  {journal} {Phys. Rev. B}\ }\textbf {\bibinfo {volume} {20}},\ \bibinfo
  {pages} {169} (\bibinfo {year} {1979})}\BibitemShut {NoStop}%
\bibitem [{\citenamefont {Mermin}\ and\ \citenamefont
  {Ho}(1976)}]{MerminHoRelation}%
  \BibitemOpen
  \bibfield  {author} {\bibinfo {author} {\bibfnamefont {N.~D.}\ \bibnamefont
  {Mermin}}\ and\ \bibinfo {author} {\bibfnamefont {T.-L.}\ \bibnamefont
  {Ho}},\ }\bibfield  {title} {\bibinfo {title} {Circulation and angular
  momentum in the $\mathrm{A}$ phase of superfluid helium-3},\ }\href
  {https://doi.org/10.1103/PhysRevLett.36.594} {\bibfield  {journal} {\bibinfo
  {journal} {Phys. Rev. Lett.}\ }\textbf {\bibinfo {volume} {36}},\ \bibinfo
  {pages} {594} (\bibinfo {year} {1976})}\BibitemShut {NoStop}%
\bibitem [{\citenamefont {Fetter}(1979)}]{FetterCoeff}%
  \BibitemOpen
  \bibfield  {author} {\bibinfo {author} {\bibfnamefont {A.~L.}\ \bibnamefont
  {Fetter}},\ }\bibfield  {title} {\bibinfo {title} {Stability of helical
  textures in $^{3}\mathrm{He}$-$\mathrm{A}$},\ }\href
  {https://doi.org/10.1103/PhysRevB.20.303} {\bibfield  {journal} {\bibinfo
  {journal} {Phys. Rev. B}\ }\textbf {\bibinfo {volume} {20}},\ \bibinfo
  {pages} {303} (\bibinfo {year} {1979})}\BibitemShut {NoStop}%
\bibitem [{\citenamefont {Thuneberg}()}]{ThunebergCoeff}%
  \BibitemOpen
  \bibfield  {author} {\bibinfo {author} {\bibfnamefont {E.}~\bibnamefont
  {Thuneberg}},\ }\href@noop {} {\bibinfo {title} {{BCS} gap function etc,
  parameters for $^3\mathrm{He}$}},\ \bibinfo {howpublished}
  {\url{https://users.aalto.fi/~thunebe1/theory/qc/bcsgap.html}},\ \bibinfo
  {note} {accessed: 2022-07-02}\BibitemShut {NoStop}%
\bibitem [{\citenamefont {Ambegaokar}\ \emph {et~al.}(1974)\citenamefont
  {Ambegaokar}, \citenamefont {deGennes},\ and\ \citenamefont
  {Rainer}}]{lboundarycondition}%
  \BibitemOpen
  \bibfield  {author} {\bibinfo {author} {\bibfnamefont {V.}~\bibnamefont
  {Ambegaokar}}, \bibinfo {author} {\bibfnamefont {P.~G.}\ \bibnamefont
  {deGennes}},\ and\ \bibinfo {author} {\bibfnamefont {D.}~\bibnamefont
  {Rainer}},\ }\bibfield  {title} {\bibinfo {title} {{Landau-Ginsburg}
  equations for an anisotropic superfluid},\ }\href
  {https://doi.org/10.1103/PhysRevA.9.2676} {\bibfield  {journal} {\bibinfo
  {journal} {Phys. Rev. A}\ }\textbf {\bibinfo {volume} {9}},\ \bibinfo {pages}
  {2676} (\bibinfo {year} {1974})}\BibitemShut {NoStop}%
\bibitem [{\citenamefont {Brinkman}\ and\ \citenamefont
  {Cross}(1978)}]{brinkmancross}%
  \BibitemOpen
  \bibfield  {author} {\bibinfo {author} {\bibfnamefont {W.~F.}\ \bibnamefont
  {Brinkman}}\ and\ \bibinfo {author} {\bibfnamefont {M.~C.}\ \bibnamefont
  {Cross}},\ }\bibfield  {title} {\bibinfo {title} {Spin and orbital dynamics
  of superfluid $^3\mathrm{He}$},\ }in\ \href@noop {} {\emph {\bibinfo
  {booktitle} {Progress in Low Temperature Physics}}},\ Vol.\ \bibinfo {volume}
  {VIIa},\ \bibinfo {editor} {edited by\ \bibinfo {editor} {\bibfnamefont
  {D.~F.}\ \bibnamefont {Brewer}}}\ (\bibinfo  {publisher} {North-Holland},\
  \bibinfo {year} {1978})\BibitemShut {NoStop}%
\bibitem [{\citenamefont {Ruutu}\ \emph {et~al.}(1997)\citenamefont {Ruutu},
  \citenamefont {Kopu}, \citenamefont {Krusius}, \citenamefont {Parts},
  \citenamefont {Pla\ifmmode~\mbox{\c{c}}\else \c{c}\fi{}ais}, \citenamefont
  {Thuneberg},\ and\ \citenamefont {Xu}}]{criticalvelocity}%
  \BibitemOpen
  \bibfield  {author} {\bibinfo {author} {\bibfnamefont {V.~M.~H.}\
  \bibnamefont {Ruutu}}, \bibinfo {author} {\bibfnamefont {J.}~\bibnamefont
  {Kopu}}, \bibinfo {author} {\bibfnamefont {M.}~\bibnamefont {Krusius}},
  \bibinfo {author} {\bibfnamefont {U.}~\bibnamefont {Parts}}, \bibinfo
  {author} {\bibfnamefont {B.}~\bibnamefont {Pla\ifmmode~\mbox{\c{c}}\else
  \c{c}\fi{}ais}}, \bibinfo {author} {\bibfnamefont {E.~V.}\ \bibnamefont
  {Thuneberg}},\ and\ \bibinfo {author} {\bibfnamefont {W.}~\bibnamefont
  {Xu}},\ }\bibfield  {title} {\bibinfo {title} {Critical velocity of vortex
  nucleation in rotating superfluid $^3${He}-$\mathrm{A}$},\ }\href
  {https://doi.org/10.1103/PhysRevLett.79.5058} {\bibfield  {journal} {\bibinfo
   {journal} {Phys. Rev. Lett.}\ }\textbf {\bibinfo {volume} {79}},\ \bibinfo
  {pages} {5058} (\bibinfo {year} {1997})}\BibitemShut {NoStop}%
\bibitem [{\citenamefont {Parts}\ \emph
  {et~al.}(1995{\natexlab{a}})\citenamefont {Parts}, \citenamefont
  {Karim\"aki}, \citenamefont {Koivuniemi}, \citenamefont {Krusius},
  \citenamefont {Ruutu}, \citenamefont {Thuneberg},\ and\ \citenamefont
  {Volovik}}]{vortexphasediagram}%
  \BibitemOpen
  \bibfield  {author} {\bibinfo {author} {\bibfnamefont {U.}~\bibnamefont
  {Parts}}, \bibinfo {author} {\bibfnamefont {J.~M.}\ \bibnamefont
  {Karim\"aki}}, \bibinfo {author} {\bibfnamefont {J.~H.}\ \bibnamefont
  {Koivuniemi}}, \bibinfo {author} {\bibfnamefont {M.}~\bibnamefont {Krusius}},
  \bibinfo {author} {\bibfnamefont {V.~M.~H.}\ \bibnamefont {Ruutu}}, \bibinfo
  {author} {\bibfnamefont {E.~V.}\ \bibnamefont {Thuneberg}},\ and\ \bibinfo
  {author} {\bibfnamefont {G.~E.}\ \bibnamefont {Volovik}},\ }\bibfield
  {title} {\bibinfo {title} {Phase diagram of vortices in superfluid
  $^3\mathrm{He}$-$\mathrm{A}$},\ }\href
  {https://doi.org/10.1103/PhysRevLett.75.3320} {\bibfield  {journal} {\bibinfo
   {journal} {Phys. Rev. Lett.}\ }\textbf {\bibinfo {volume} {75}},\ \bibinfo
  {pages} {3320} (\bibinfo {year} {1995}{\natexlab{a}})}\BibitemShut {NoStop}%
\bibitem [{\citenamefont {Blaauwgeers}\ \emph {et~al.}(2008)\citenamefont
  {Blaauwgeers}, \citenamefont {Eltsov}, \citenamefont {Krusius}, \citenamefont
  {Ruohio},\ and\ \citenamefont {Schanen}}]{quantizedvorticity}%
  \BibitemOpen
  \bibfield  {author} {\bibinfo {author} {\bibfnamefont {R.}~\bibnamefont
  {Blaauwgeers}}, \bibinfo {author} {\bibfnamefont {V.}~\bibnamefont {Eltsov}},
  \bibinfo {author} {\bibfnamefont {M.}~\bibnamefont {Krusius}}, \bibinfo
  {author} {\bibfnamefont {J.}~\bibnamefont {Ruohio}},\ and\ \bibinfo {author}
  {\bibfnamefont {R.}~\bibnamefont {Schanen}},\ }\bibinfo {title} {Quantized
  vorticity in superfluid $^3\mathrm{He}$-$\mathrm{A}$: Structure and
  dynamics}\ (\bibinfo {year} {2008})\ pp.\ \bibinfo {pages}
  {399--420}\BibitemShut {NoStop}%
\bibitem [{\citenamefont {Takagi}(2005)}]{TakagiNMR}%
  \BibitemOpen
  \bibfield  {author} {\bibinfo {author} {\bibfnamefont {T.}~\bibnamefont
  {Takagi}},\ }\bibfield  {title} {\bibinfo {title} {Textures and {NMR} spectra
  of a superfluid $^3\mathrm{He}$-$\mathrm{A}$ phase confined in rotating
  narrow cylinders},\ }\href
  {https://doi.org/https://doi.org/10.1016/j.jpcs.2005.05.010} {\bibfield
  {journal} {\bibinfo  {journal} {Journal of Physics and Chemistry of Solids}\
  }\textbf {\bibinfo {volume} {66}},\ \bibinfo {pages} {1355} (\bibinfo {year}
  {2005})},\ \bibinfo {note} {proceedings of the ISSP International Symposium
  (ISSP-9)on Quantum Condensed System}\BibitemShut {NoStop}%
\bibitem [{\citenamefont {Anderson}\ and\ \citenamefont
  {Toulouse}(1977)}]{ATC1}%
  \BibitemOpen
  \bibfield  {author} {\bibinfo {author} {\bibfnamefont {P.~W.}\ \bibnamefont
  {Anderson}}\ and\ \bibinfo {author} {\bibfnamefont {G.}~\bibnamefont
  {Toulouse}},\ }\bibfield  {title} {\bibinfo {title} {Phase slippage without
  vortex cores: Vortex textures in superfluid $^{3}\mathrm{He}$},\ }\href
  {https://doi.org/10.1103/PhysRevLett.38.508} {\bibfield  {journal} {\bibinfo
  {journal} {Phys. Rev. Lett.}\ }\textbf {\bibinfo {volume} {38}},\ \bibinfo
  {pages} {508} (\bibinfo {year} {1977})}\BibitemShut {NoStop}%
\bibitem [{\citenamefont {Chechetkin}(1976)}]{ATC2}%
  \BibitemOpen
  \bibfield  {author} {\bibinfo {author} {\bibfnamefont {V.~R.}\ \bibnamefont
  {Chechetkin}},\ }\bibfield  {title} {\bibinfo {title} {Types of vortex
  solutions in superfluid $\mathrm{He}^3$},\ }\href@noop {} {\bibfield
  {journal} {\bibinfo  {journal} {Sov. Phys. JETP}\ }\textbf {\bibinfo {volume}
  {44}},\ \bibinfo {pages} {766} (\bibinfo {year} {1976})}\BibitemShut
  {NoStop}%
\bibitem [{\citenamefont {Lounasmaa}\ and\ \citenamefont
  {Thuneberg}(1999)}]{ThunebergPNAS}%
  \BibitemOpen
  \bibfield  {author} {\bibinfo {author} {\bibfnamefont {O.~V.}\ \bibnamefont
  {Lounasmaa}}\ and\ \bibinfo {author} {\bibfnamefont {E.}~\bibnamefont
  {Thuneberg}},\ }\bibfield  {title} {\bibinfo {title} {Vortices in rotating
  superfluid $^3\mathrm{He}$},\ }\href
  {https://doi.org/10.1073/pnas.96.14.7760} {\bibfield  {journal} {\bibinfo
  {journal} {Proceedings of the National Academy of Sciences}\ }\textbf
  {\bibinfo {volume} {96}},\ \bibinfo {pages} {7760} (\bibinfo {year}
  {1999})},\ \Eprint
  {https://arxiv.org/abs/https://www.pnas.org/doi/pdf/10.1073/pnas.96.14.7760}
  {https://www.pnas.org/doi/pdf/10.1073/pnas.96.14.7760} \BibitemShut {NoStop}%
\bibitem [{\citenamefont {Eltsov}\ and\ \citenamefont
  {Krusius}(2000)}]{topologicaldefects}%
  \BibitemOpen
  \bibfield  {author} {\bibinfo {author} {\bibfnamefont {V.}~\bibnamefont
  {Eltsov}}\ and\ \bibinfo {author} {\bibfnamefont {M.}~\bibnamefont
  {Krusius}},\ }\bibfield  {title} {\bibinfo {title} {Topological defects in
  $^3\mathrm{He}$ superfluids},\ }\href
  {https://doi.org/10.1007/978-94-011-4106-2_15} {\ \textbf {\bibinfo {volume}
  {549}},\ \bibinfo {pages} {325} (\bibinfo {year} {2000})}\BibitemShut
  {NoStop}%
\bibitem [{\citenamefont {Parts}\ \emph
  {et~al.}(1995{\natexlab{b}})\citenamefont {Parts}, \citenamefont {Ruutu},
  \citenamefont {Koivuniemi}, \citenamefont {Krusius}, \citenamefont
  {Thuneberg},\ and\ \citenamefont {Volovik}}]{SheetMeasurements}%
  \BibitemOpen
  \bibfield  {author} {\bibinfo {author} {\bibfnamefont {U.}~\bibnamefont
  {Parts}}, \bibinfo {author} {\bibfnamefont {V.}~\bibnamefont {Ruutu}},
  \bibinfo {author} {\bibfnamefont {J.}~\bibnamefont {Koivuniemi}}, \bibinfo
  {author} {\bibfnamefont {M.}~\bibnamefont {Krusius}}, \bibinfo {author}
  {\bibfnamefont {E.}~\bibnamefont {Thuneberg}},\ and\ \bibinfo {author}
  {\bibfnamefont {G.}~\bibnamefont {Volovik}},\ }\bibfield  {title} {\bibinfo
  {title} {Measurements on the vortex sheet in rotating superfluid
  $^3\mathrm{He}$-$\mathrm{A}$},\ }\href
  {https://doi.org/https://doi.org/10.1016/0921-4526(94)01116-I} {\bibfield
  {journal} {\bibinfo  {journal} {Physica B: Condensed Matter}\ }\textbf
  {\bibinfo {volume} {210}},\ \bibinfo {pages} {311} (\bibinfo {year}
  {1995}{\natexlab{b}})},\ \bibinfo {note} {vortices, Interfaces and Mesoscopic
  Phenomena in Quantum Systems}\BibitemShut {NoStop}%
\bibitem [{\citenamefont {Thuneberg}(1995)}]{ThunebergSheet}%
  \BibitemOpen
  \bibfield  {author} {\bibinfo {author} {\bibfnamefont {E.}~\bibnamefont
  {Thuneberg}},\ }\bibfield  {title} {\bibinfo {title} {Introduction to the
  vortex sheet of superfluid $^3\mathrm{He}$-$\mathrm{A}$},\ }\href
  {https://doi.org/https://doi.org/10.1016/0921-4526(94)01114-G} {\bibfield
  {journal} {\bibinfo  {journal} {Physica B: Condensed Matter}\ }\textbf
  {\bibinfo {volume} {210}},\ \bibinfo {pages} {287} (\bibinfo {year}
  {1995})},\ \bibinfo {note} {vortices, Interfaces and Mesoscopic Phenomena in
  Quantum Systems}\BibitemShut {NoStop}%
\bibitem [{\citenamefont {Leggett}(1972)}]{SBSOS1}%
  \BibitemOpen
  \bibfield  {author} {\bibinfo {author} {\bibfnamefont {A.~J.}\ \bibnamefont
  {Leggett}},\ }\bibfield  {title} {\bibinfo {title} {Interpretation of recent
  results on $\mathrm{He}^3$ below 3 m{K}: A new liquid phase?},\ }\href
  {https://doi.org/10.1103/PhysRevLett.29.1227} {\bibfield  {journal} {\bibinfo
   {journal} {Phys. Rev. Lett.}\ }\textbf {\bibinfo {volume} {29}},\ \bibinfo
  {pages} {1227} (\bibinfo {year} {1972})}\BibitemShut {NoStop}%
\bibitem [{\citenamefont {Leggett}(1973)}]{SBSOS2}%
  \BibitemOpen
  \bibfield  {author} {\bibinfo {author} {\bibfnamefont {A.~J.}\ \bibnamefont
  {Leggett}},\ }\bibfield  {title} {\bibinfo {title} {Microscopic theory of
  {NMR} in an anisotropic superfluid ($^{3}\mathrm{He}$-$\mathrm{A}$)},\ }\href
  {https://doi.org/10.1103/PhysRevLett.31.352} {\bibfield  {journal} {\bibinfo
  {journal} {Phys. Rev. Lett.}\ }\textbf {\bibinfo {volume} {31}},\ \bibinfo
  {pages} {352} (\bibinfo {year} {1973})}\BibitemShut {NoStop}%
\bibitem [{\citenamefont {Ruutu}\ \emph {et~al.}(1996)\citenamefont {Ruutu},
  \citenamefont {Parts},\ and\ \citenamefont {Krusius}}]{NMRsignatures}%
  \BibitemOpen
  \bibfield  {author} {\bibinfo {author} {\bibfnamefont {V.~M.~H.}\
  \bibnamefont {Ruutu}}, \bibinfo {author} {\bibfnamefont {U.}~\bibnamefont
  {Parts}},\ and\ \bibinfo {author} {\bibfnamefont {M.}~\bibnamefont
  {Krusius}},\ }\bibfield  {title} {\bibinfo {title} {{NMR} signatures of
  topological objects in rotating superfluid $^3${H}e-{A}},\ }\href
  {https://doi.org/10.1007/BF00754793} {\bibfield  {journal} {\bibinfo
  {journal} {J. Low Temp. Phys.}\ }\textbf {\bibinfo {volume} {103}},\ \bibinfo
  {pages} {331} (\bibinfo {year} {1996})}\BibitemShut {NoStop}%
\bibitem [{\citenamefont {Hänninen}\ and\ \citenamefont
  {Thuneberg}(2001)}]{ThunebergNMR}%
  \BibitemOpen
  \bibfield  {author} {\bibinfo {author} {\bibfnamefont {R.}~\bibnamefont
  {Hänninen}}\ and\ \bibinfo {author} {\bibfnamefont {E.}~\bibnamefont
  {Thuneberg}},\ }\bibfield  {title} {\bibinfo {title} {Calculation of {NMR}
  properties of solitons in superfluid $^3\mathrm{He}$-$\mathrm{A}$},\ }\href
  {https://doi.org/10.1103/PhysRevB.68.094504} {\bibfield  {journal} {\bibinfo
  {journal} {Physical Review B}\ }\textbf {\bibinfo {volume} {68}} (\bibinfo
  {year} {2001})}\BibitemShut {NoStop}%
\bibitem [{\citenamefont {Volovik}(1988)}]{volovik}%
  \BibitemOpen
  \bibfield  {author} {\bibinfo {author} {\bibfnamefont {G.~E.}\ \bibnamefont
  {Volovik}},\ }\bibfield  {title} {\bibinfo {title} {Effect of zero-charge
  phenomena on nonsingular vortex in $^3\mathrm{He}$-$\mathrm{A}$},\
  }\href@noop {} {\bibfield  {journal} {\bibinfo  {journal} {Pis'ma Zh. Eksp.
  Teor. Fiz.}\ }\textbf {\bibinfo {volume} {47}},\ \bibinfo {pages} {46}
  (\bibinfo {year} {1988})}\BibitemShut {NoStop}%
\bibitem [{\citenamefont {Blaauwgeers}\ \emph {et~al.}(2000)\citenamefont
  {Blaauwgeers}, \citenamefont {Eltsov}, \citenamefont {Krusius}, \citenamefont
  {Ruohio}, \citenamefont {Schanen},\ and\ \citenamefont
  {Volovik}}]{DQVNature}%
  \BibitemOpen
  \bibfield  {author} {\bibinfo {author} {\bibfnamefont {R.}~\bibnamefont
  {Blaauwgeers}}, \bibinfo {author} {\bibfnamefont {V.}~\bibnamefont {Eltsov}},
  \bibinfo {author} {\bibfnamefont {M.}~\bibnamefont {Krusius}}, \bibinfo
  {author} {\bibfnamefont {J.}~\bibnamefont {Ruohio}}, \bibinfo {author}
  {\bibfnamefont {R.}~\bibnamefont {Schanen}},\ and\ \bibinfo {author}
  {\bibfnamefont {G.}~\bibnamefont {Volovik}},\ }\bibfield  {title} {\bibinfo
  {title} {Double-quantum vortex in superfluid $^3\mathrm{He}$-$\mathrm{A}$},\
  }\href {https://doi.org/10.1038/35006583} {\bibfield  {journal} {\bibinfo
  {journal} {Nature}\ }\textbf {\bibinfo {volume} {404}},\ \bibinfo {pages}
  {471} (\bibinfo {year} {2000})}\BibitemShut {NoStop}%
\end{thebibliography}%

\end{document}